%% file: main.tex
\documentclass{article}

\usepackage{arxiv}

\usepackage[utf8]{inputenc} %
\usepackage[T1]{fontenc}    %
\usepackage{hyperref}       %
\usepackage{url}            %
\usepackage{booktabs}       %
\usepackage{amsfonts}       %
\usepackage{nicefrac}       %
\usepackage{microtype}      %
\usepackage{lipsum}
\usepackage{graphicx}
\usepackage{float}
\usepackage{subcaption}
\usepackage{commath}
\usepackage{placeins}
\usepackage{amsmath}
\usepackage{relsize}
\usepackage[toc,page]{appendix}
\usepackage{tikz}
\usepackage{cite}
\usepackage{todonotes}

\title{MAD-VAE: Manifold Awareness Defense Variational Autoencoder}

\author{
 Frederick Morlock \\
  New York University Shanghai\\
  \texttt{fm1391@nyu.edu} \\
   \And
  Dingsu Wang\\
  New York University Shanghai\\
  \texttt{dw1920@nyu.edu} \\
}

\begin{document}
\maketitle

\begin{abstract}
Although deep generative models such as Defense-GAN \cite{samangouei2018defense} and Defense-VAE \cite{defense-vae18} have made significant progress in terms of adversarial defenses of image classification neural networks, several methods have been found to circumvent these defenses \cite{jalal2017robust} \cite{athalye2018obfuscated}. Based on Defense-VAE, in our research we introduce several methods to improve the robustness of defense models. The methods introduced in this paper are straight forward yet show promise over the vanilla Defense-VAE. With extensive experiments on MNIST \cite{lecun-mnisthandwrittendigit-2010} data set, we have demonstrated the effectiveness of our algorithms against different attacks. Our experiments also include attacks on the latent space \cite{jalal2017robust} of the defensive model. We also discuss the applicability of existing adversarial latent space attacks as they may have a significant flaw.
\end{abstract}


\section{Introduction}
\footnotetext[1]{Our code is available on Github at \url{https://github.com/Derek-Wds/MAD-VAE}}%
With the development of deep learning in recent years, research about image classification have made significant breakthroughs. Models such as ResNet \cite{xie2017aggregated} and VGG \cite{simonyan2014very} as well as their variants have achieved amazing accuracy on image classification tasks. Despite their high accuracy on image classification, neural network based models are often vulnerable to misclassification given a well-crafted input by an adversary. The differences between these \textit{adversarial examples} and original data are often imperceptible to humans, but result in misclassification by the model \cite{goodfellow2014explaining, szegedy2013intriguing}. 

Recent research has considered two main attack methods: White-Box attack and Black-Box attack. Under the White-Box attack scenario, the attackers have access to, at the very least, the classification model as well as the pretrained parameters. Under the Black-Box attack situation, the attackers do not have any information about the classification model \cite{xu2019adversarial}.

Along with attacks, various defenses mechanisms have also been proposed in recent years. The defenses methods can mainly be divided into three groups \cite{xu2019adversarial}: (1) obfuscating the gradient of the classification model \cite{defense-vae18, samangouei2018defense, goodfellow2014explaining, kurakin2016adversarial, dong2017boosting}, (2) robust optimization such as such as training the classification model with adversarial examples \cite{goodfellow2014explaining} and (3) adversary detection before feeding images to the classification model \cite{grosse2017statistical, Xu_2018}. 

In our research, we focus on the first kind of defense mechanism, gradient obfuscation, and improve upon the Defense-VAE \cite{defense-vae18} model. The idea of Defense-VAE is to use VAE's ability to learn the distribution of input data in order to transfer the potentially adversarial input data back to the data manifold. In the original VAE framework \cite{kingma2013auto}, the model consists of two parts: an encoder $E$ which maps the input data onto a latent distribution, and a decoder $D$ which tries to reconstruct the input data based on the variable $z$ sampled from the latent distribution. Defense-VAE leverages this idea in their defense model and proves that it is a robust way to prevent against attacks. Unfortunately, although generative defense models such as Defense-GAN \cite{samangouei2018defense} and Defense-VAE \cite{defense-vae18} have achieved decent results on prevent certain attacks, they have been shown to still be vulnerable to carefully crafted adversarial examples that attack the latent space of the generative model instead of the classifier directly \cite{jalal2017robust}.

The following paper consists of following parts: for theoretical background and related works, please see Appendix \ref{related}; in Section \ref{proposed}, we will introduce the research work we are focusing on; Section \ref{experiments} will be about the experiments and results while Section \ref{conclusion} discusses future works and concludes the paper. There are additional plots and tables in the remainder of the Appendix that were not included in the main body of the paper.

\section{Proposed MAD-VAE} \label{proposed}
We propose modifying the training procedure and introduce several additional loss functions, which help Defense-VAE to be more robust against adversarial attack \cite{jalal2017robust}. We call our model Manifold Awareness Defense-VAE because our new training steps attempt to capture the topological structure of the data manifold as well as the latent variables. In the following subsections, we will first talk about the motivation of our research and then describe our methods.

\subsection{Motivation}
In order to further mitigate the problem of adversarial attack, we formulate our objective in the following way: we want the encoded $z$ to represent the input data well while having the decoded $G(z)$ be correctly classified. Since $z_i$ follows a certain distribution for each class $i$, we want the points $z_i^j$ sampled from this distribution to be close to each other while maintaining that this cluster is well separated from clusters of other classes. Ideally, our model should be able to encode the adversarial data $X'$, where the benign $X$ data has label $i$, onto or close to the distribution $z_i$. Therefore, when we sample from the latent distribution and try to find $z_i^j$ that decodes to similar data, not only will it be less likely to decode data that lies in the other clusters but also the correct classification can be ensured.

\subsection{Specification}
We introduce three loss functions that increase the robustness of Defense-VAE through improving the awareness of the underlying topology of the data manifold. This is a realization of the goals mentioned in the previous section.

\subsubsection{Classification Loss}
When the classifier gets an input data $X$, it will extract certainly features from $X$ and make classification based on them. Since our goal is to make sure that the output of our model can be correctly classified, one straight forward way to improve the model is to having a classification loss on the output of the VAE $G(z)$. This idea is from GAN (Generative Adversarial Networks) \cite{goodfellow2014generative} where we have a generative model $G$ to generate data and a discriminator $D$ to make sure the generated data is what we want. A similar idea was proposed by Jang et al. \cite{jang2019need} for improving the topological awareness of generative defense models. Specifically, in our case, we have the VAE as our generative model and the pretrained classifier to be our discriminator model.

During training of the VAE, we fix the parameters of the classifier and feed the output of VAE $G(z)$ to the classifier to get a classification loss $L_c$. The new loss function of our model is as follows:
\begin{align*}
    L = L_r + \beta \cdot L_{kl} + \alpha \cdot L_c
\end{align*}
where $L_r$ is the reconstruction loss and $L_{kl}$ is the KL-divergence loss in the \textit{ELBO} equation. $\alpha$ and $\beta$ are two hyperparameters of the weight on different loss.

By doing so, we hypothesize that the classification loss can help to coerce the VAE model to generate features that are sensitive to the classifier, and at the same time, force the latent $z$ to lie in the cluster of $z_i$ for each class $i$.

\subsubsection{Proximity and Distance Loss}

According to Zhao et al. \cite{zhao2017deeper}, due to the nature of MNIST, it would be hard for a model like our own to learn a clear separation between classes and as a result the latent distributions of $z$ will certainly mix with each other. In order to solve this problem, we decide to add another explicit loss to constrain the latent variables. Similar to the idea from the research conducted by Mustafa et al. \cite{Mustafa_2019_ICCV}, we try to learn the center $c_i$ of the cluster of $z_i$ for each class $i$ and, at the same time, keep the clusters away from each other. The proposed loss of function can be described as follows:
\begin{align*}
    L_p & = ||z_i - c_i||_2 \\
    L_d & = \frac{1}{k-1} \cdot \sum_{j\neq i} \left(||z_i - c_j||_2 + ||c_i - c_j||_2\right) \\
    L_{pd} & = \sum_i\left[\alpha \cdot L_p - \sigma \cdot L_d \right]
\end{align*}
where $k$ is total number of labels in the dataset and the distances are computed by the euclidean distances. $\alpha$ and $\sigma$ are the weights between the proximity loss and distance loss respectively. Therefore, the loss function for our VAE model becomes:
\begin{align*}
    L = L_r + \beta \cdot L_{kl} + L_{pd}
\end{align*}
where the variables are the same as those above. The idea of this loss function is to force the latent variable $z_i$ with the same label $i$ to gather together towards the center $c_i$ while make sure $c_i$ and $z_i$ is farther away from the center of cluster of $z_j$ where $j \neq i$. Through the adversarial training on our defensive model, we are pushing the decision boundaries between each cluster to better separate the classes. 

\subsubsection{Combined Loss}
Since we have discussed two new loss functions for generative defense models in the subsections above, it would be natural to think of a way to combine them together in order to utilize both their advantages. A naive approach to the aggregation of the two defense loss functions would be to design a weighted sum. To do so, we use the combined loss below for our VAE model:
\begin{align*}
    L = L_r + \beta \cdot L_{kl} + \gamma \cdot L_c + L_{pd}
\end{align*}
where $L_r$ is the reconstruction loss and $L_{kl}$ is the KL-divergence loss in the $ELBO$ equation. $\beta$ and $\gamma$ are two hyperparameters controlling the weight assigned to the different loss functions.

\section{Experiments} \label{experiments}
We compare our methods against the original Defense-VAE model, however, it may be noted that our model does not use any of the Defense-VAE's code base. Through our reproduction of Defense-VAE, we were able to match the performance detailed in the Defense-VAE paper \cite{defense-vae18}, therefore validating its functional equivalence.

We evaluate our defensive model under the FGSM, Rand-FGSM and CW white-box attacks as well as three other attack methods not included in our training process: PGD \cite{aleks2017deep}, momentum iterative FGSM \cite{dong2017boosting} and single pixel attack \cite{narodytska2016simple}. We also evaluate MAD-VAE and Defense-VAE under the Overpowered Attack \cite{jalal2017robust}. Although the Overpowered Attack is technically a white-box attack, it is a \textit{latent space} attack and therefore it is not reasonable to compare it against other white-box attacks. Instead, the Overpowered Attack gets its own section, after Black-Box Attacks.

Our classifier and MAD-VAE implementation are written using the PyTorch \cite{paszke2019pytorch} library and we use the open source package AdverTorch \cite{ding2019advertorch} for performing various attack methods. The Overpowered Attack is also written in PyTorch and is largely based off of the code provided by Jalal et al. \cite{jalal2017robust}.

We use the MNIST dataset \cite{lecun-mnisthandwrittendigit-2010} for our experiments, which contains 60,000 training images and 10,000 testing images. We generate a large, combined training dataset using three attack methods: FGSM, Rand-FGSM and CW with three different parameter settings for each one of them. We then use another parameter setting to generate the validation dataset to evaluate the performance of our models, and we use the testing data set to get our test dataset for all the six attack algorithms mentioned above. Since the attack generating process is stochastic, it is rare for non-empty intersections of different dataset to occur. 

Due to time limitations and the computational complexity of the models, we have not been able to fully reproduce the experiments conducted in Defense-VAE \cite{defense-vae18} and Defense-GAN \cite{samangouei2018defense}. Our defense model, however, remains based on the architecture of Defense-VAE and is able to match the Defense-VAE \cite{defense-vae18} accuracy without any of our additional loss functions. The structure of MAD-VAE can be found in Table \ref{table:defense-vae} in the Appendix. In our experiments, we choose the classifier MagNet \cite{meng2017magnet} to be our classification model. MagNet's model structure can be found in Table \ref{table:classifier} in the Appendix. 

The original Defense-VAE model converges after $5$ epochs while our model, with the new loss functions the proximity and distance loss variant converges after $5$ epochs, while the other variants converge after $10$ epochs. In order to achieve the best performance, we validated our model on a validation data set for different parameter settings. From our experiments on the validation set, we chose the weight of our loss functions as: $\alpha=0.01$, $\sigma=0.00001$, $\beta=0.1$ and $\gamma=0.1$. The other parameters setting are kept the same as Defense-VAE paper. We use the Adam and SGD optimizers for our gradient descent with a exponential learning rate scheduler.

\subsection{Results on White-Box Attack}
In this section, we will present the results on White-Box attack using six different attack methods: FGSM, Rand-FGSM, CW, MI-FGSM, PGD and Single Pixel. All of the adversarial images are generated using the AdverTorch \cite{ding2019advertorch} package with default parameters. The results are shown in Table \ref{table:whitebox-result}.

\input{whitebox_table}

We note that the attacks are successfully fooling the classifier except for the Single Pixel attack. Table \ref{table:whitebox-result} demonstrated that our methods make the original Defense-VAE more robust under all the six attacks. Additionally the model utilizing Classification loss outperformed the others on most of the situations with the Combined loss outperforming under the other attacks. Unlike what was does in the Defense-VAE paper, we did not fine-tune the classifier based on the output of out model. Through our topologically aware training, we consistently outperform Defense-VAE on the tested white-box attacks.

\subsection{Results on Black-Box Attack}
In this section, we present the results of our experiments of MAD-VAE compared with the vanilla Defense-VAE on the FGSM Black-Box attack. As what is mentioned above, attackers will have no access to the classifier information under the Black-Box attack scenario. They thus training a substitute model for generating attacks. The detail of substitute models can be found in  Table \ref{table:black-box}.

In Table \ref{table:blackbox-result}, we present our classification results for different substitute classification model from the Defense-VAE paper \cite{defense-vae18}. For all the experiments, we set the FGSM parameter $\epsilon$ to be $0.3$ which is the default value given by the original paper. From the table, we can spot that the Black-Box FGSM attack can reduce the classification accuracy of different classifiers to only $5$ percent. All models, including Defense-VAE, perform well on the Black-Box FGSM attacks. However, the Classification loss model and Proximity \& Distance model outperform Defense-VAE by several percentage points.

\input{blackbox_table}

\subsection{Results under the Overpowered Attack}
The Overpowered Attack \cite{jalal2017robust} is a powerful adversarial attack that targets the latent space of generative models. In the paper by Jalal et al., the authors use the Overpowered Attack to reduce the accuracy of Defense-GAN to 3\%. Defense-VAE is not only supposed to be a faster alternative to the online-optimization of Defense-GAN, but also provides a more robust defense generative model. Because there is some noise induced by the generative defense models, attacks with different noise robustness parameters were performed both in Jalal et al. experiments and our experiments. Because this attack is a latent space attack, the classifier cannot be directly targeted to yield a baseline accuracy, hence we trained a standard  ``Identity'' VAE on MNIST.

We measured the performance of the defensive models under two metrics. In the paper by Jalal et al., they measured the accuracy of the classifier to be the number of digits such that the classifier correctly classified the digit under all noise robustness parameters, referred to as ``Original Accuracy'' in our results. On the other hand, we also measured the accuracy of the classifier on the combined pool of adversarial images, regardless of the noise robustness parameters, which we refer to as ``Accuracy''. The results from the Overpowered Attack experiments can be found in Table \ref{table:op-result}

\input{OP_table}

From Table \ref{table:op-result} we can see that under both our accuracy and the original accuracy metrics, the vanilla Defense-VAE outperformed all MAD-VAE variants, although the Proximity \& Distance variant came within 1\% of the accuracy of the vanilla Defense-VAE. These results are initially very startling due to the decrease in adversarial robustness of the MAD-VAE models in comparison to the Defense-VAE model – we expected that the MAD-VAE accuracy would either increase or remain the same relative to Defense-VAE's accuracy. The intuition for this is seeded in Section \ref{sec:clustering-op} in which we plot the distribution of these adversarial data points, both in latent and ambient space.

\subsection{Clustering for White-Box Attacks}
In this section, we will show the clustering of the adversarial examples, latent variables, and VAE output images using the Uniform Manifold Approximation and Projection (UMAP) \cite{2018arXivUMAP} dimensional reduction technique. Clustering of VAE output images can give us a direct sense of the classification result, where images with same label tend to be close to each other. Similarly, since we are trying to cluster variables in latent space, it is helpful to use clustering figures to help us evaluate the results. Since we have four models in total, in the following plots we will denote vanilla Defense-VAE as $A$, Defense-VAE with combined loss as $B$, Defense-VAE with proximity and distance loss as $C$, and Defense-VAE with classification loss as $D$.   

Since the experiments in the previous section show the strong performance of our MAD-VAE model endowed with the classification loss functions, due to the limited space we decided to only show the clustering plot of that model in this section. Additional plots for other defensive models and adversarial attacks can be found in the appendix.

\begin{figure}[H]
    \centering
    \includegraphics[width=.45\linewidth]{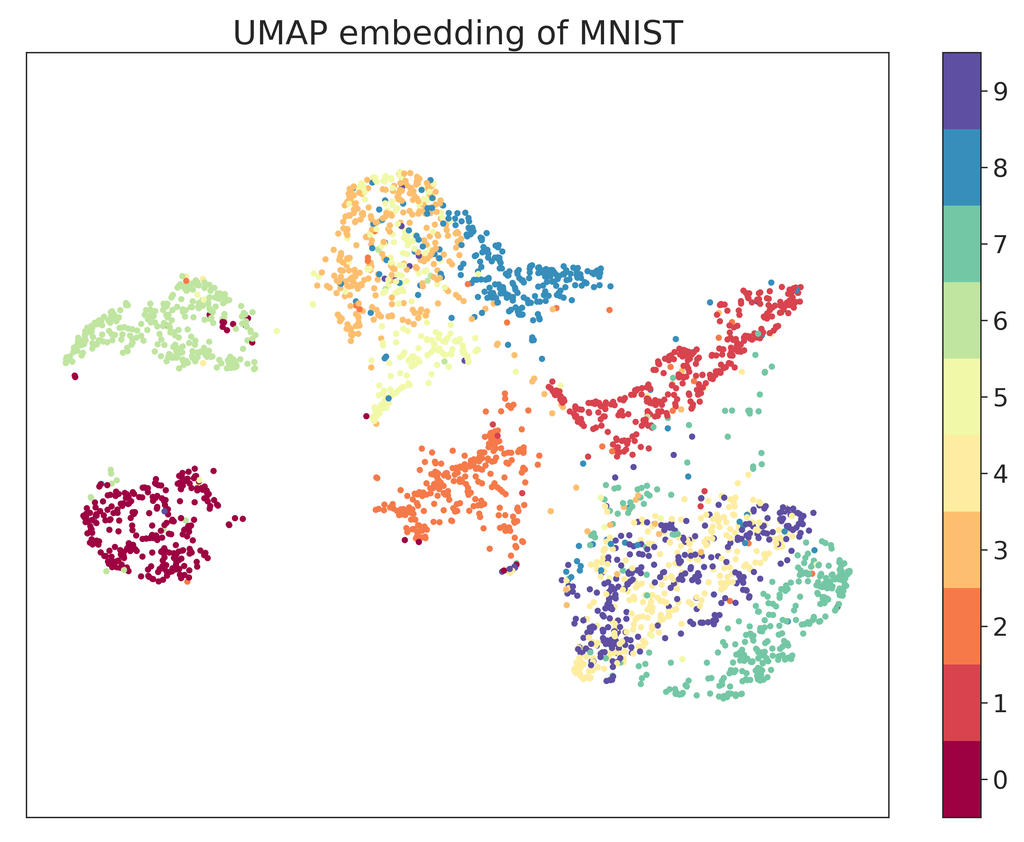}
    \caption{Clustering plots of the benign test dataset.}
    \label{fig:benign-data}
\end{figure}

Figure \ref{fig:benign-data} is the clustering of original MNIST test dataset, from which we are able to see that images of digit $6$ and $0$ are well separated from others although there have some points have similar characteristics and lie in the wrong clusters. The images of digit $9$ and $4$ are mixed with each other due to the similarities of certain features in the images, while other clusters are close to each other but still well separated. 

\begin{figure}[H]
\begin{subfigure}{.33\textwidth}
  \centering
  \includegraphics[width=\linewidth]{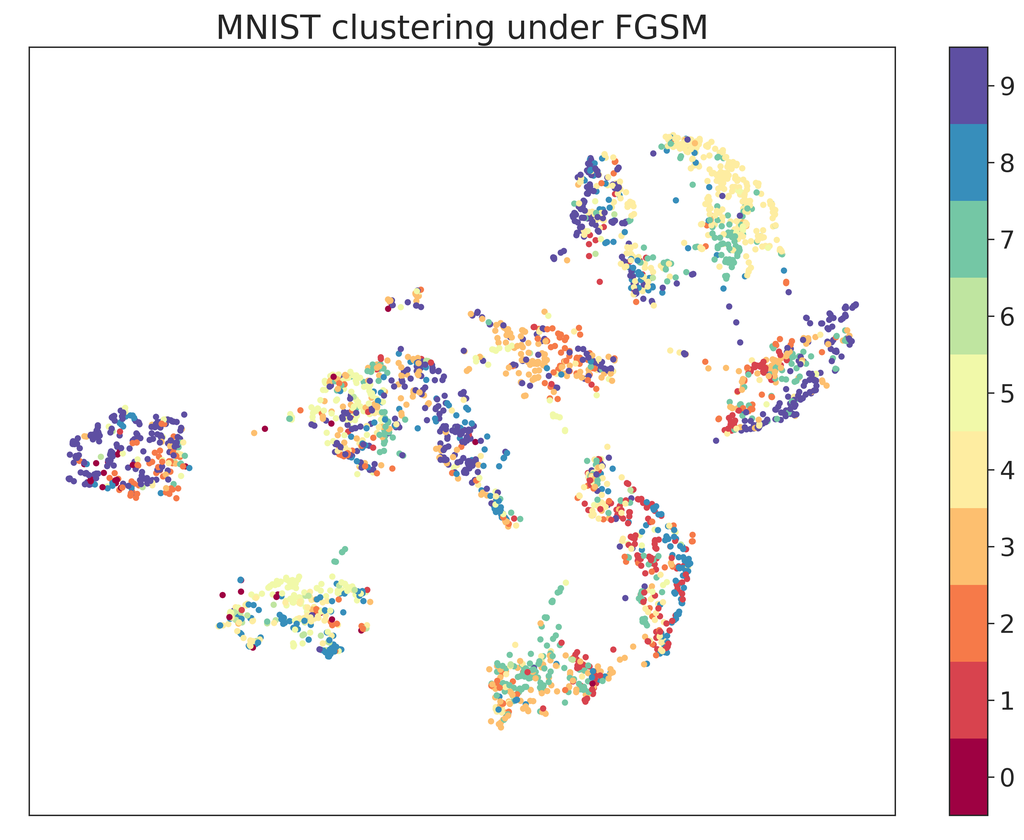}
  \caption{FGSM attack.}
  \label{fig:fgsm-clustering}
\end{subfigure}%
\begin{subfigure}{.33\textwidth}
  \centering
  \includegraphics[width=\linewidth]{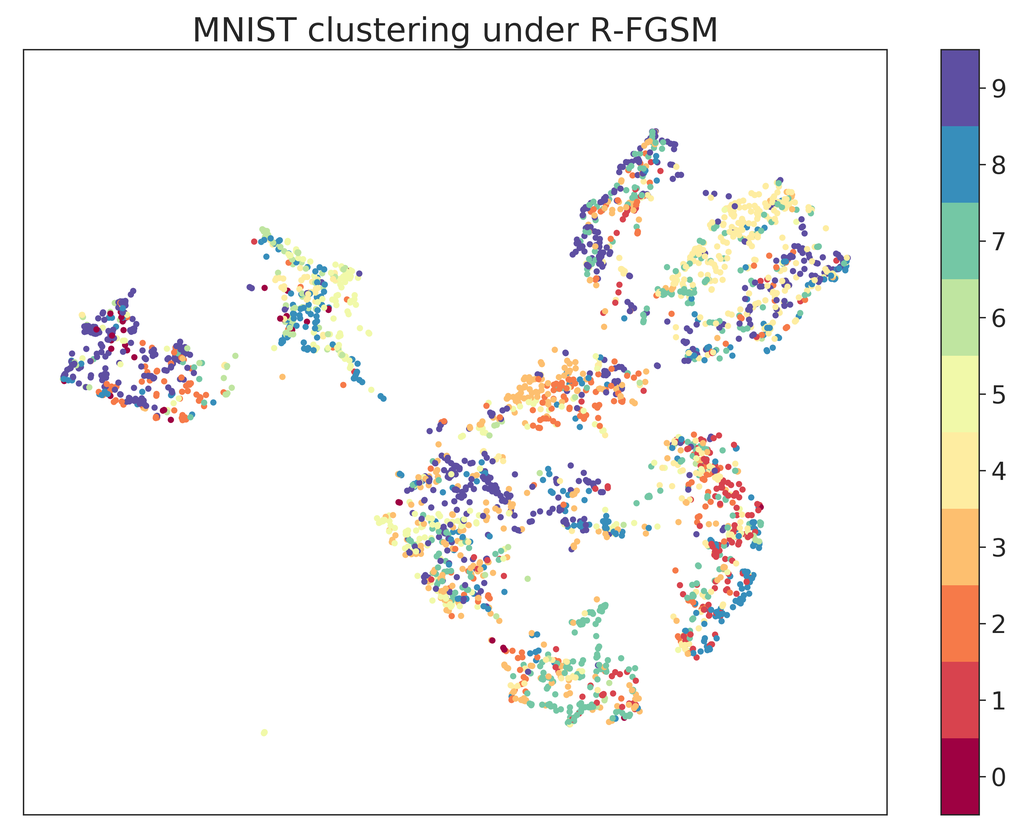}
  \caption{Rand-FGSM attacks}
  \label{fig:rfgsm-clustering}
\end{subfigure}
\begin{subfigure}{.33\textwidth}
  \centering
  \includegraphics[width=\linewidth]{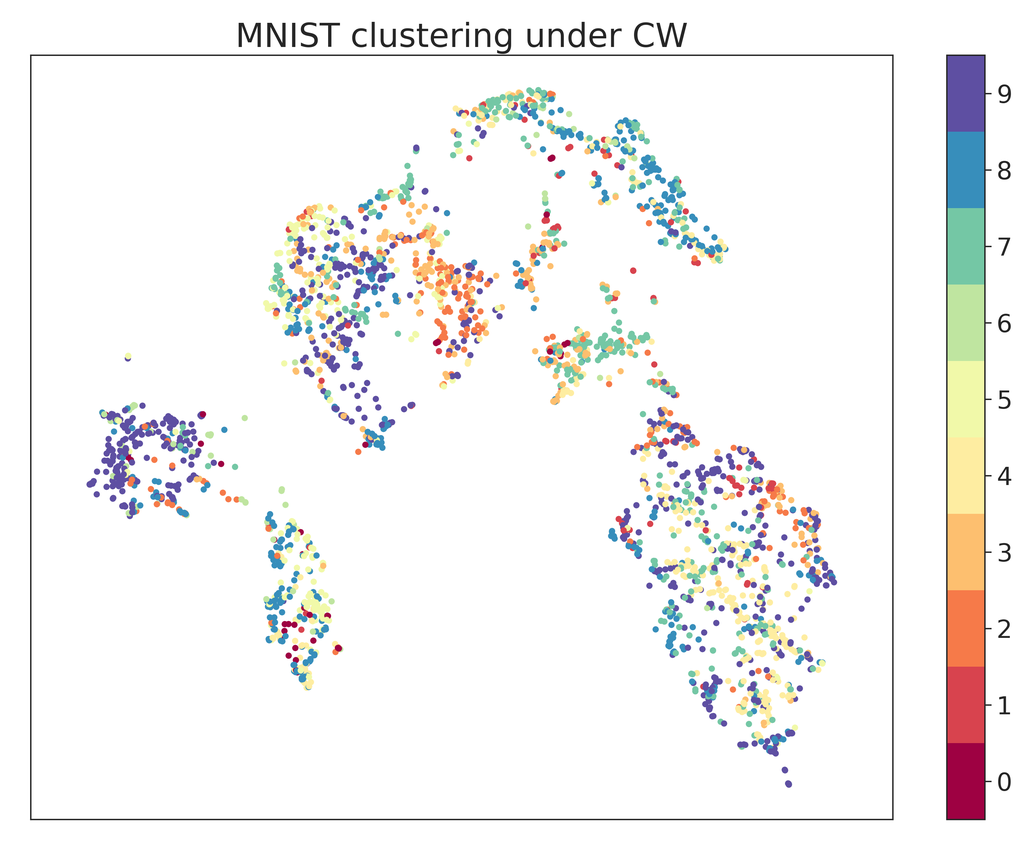}
  \caption{CW attack.}
  \label{fig:cw-clustering}
\end{subfigure}
\caption{Clustering plots for adversarial examples}
\label{fig:clustering}
\end{figure}

Figure \ref{fig:clustering} is the MNIST clustering of data and predicted labels generated from three attacks: FGSM, Rand-FGSM and CW. From the figure, we can still see clustering of the data, however, the clustering does not correspond to classes like in \ref{fig:benign-data}.

In the following two groups of plots, we will show the clustering of the output of our Defense-VAE model with the classification loss function as well as the latent variables $z$. Since our goal is to achieve data manifold awareness during training, we are hoping to spot clustering in our data and latent variables.

\begin{figure}[H]
\begin{subfigure}{.33\textwidth}
  \centering
  \includegraphics[width=1.\linewidth]{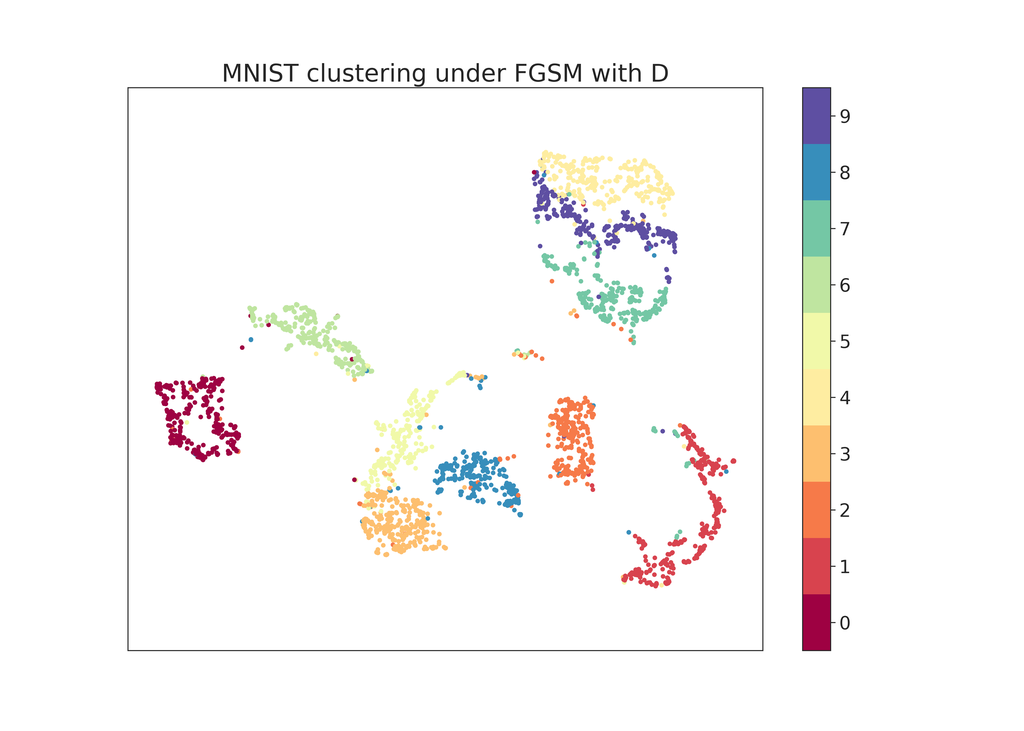}
  \caption{FGSM attack.}
  \label{fig:fgsm-output-clustering}
\end{subfigure}%
\begin{subfigure}{.33\textwidth}
  \centering
  \includegraphics[width=1.\linewidth]{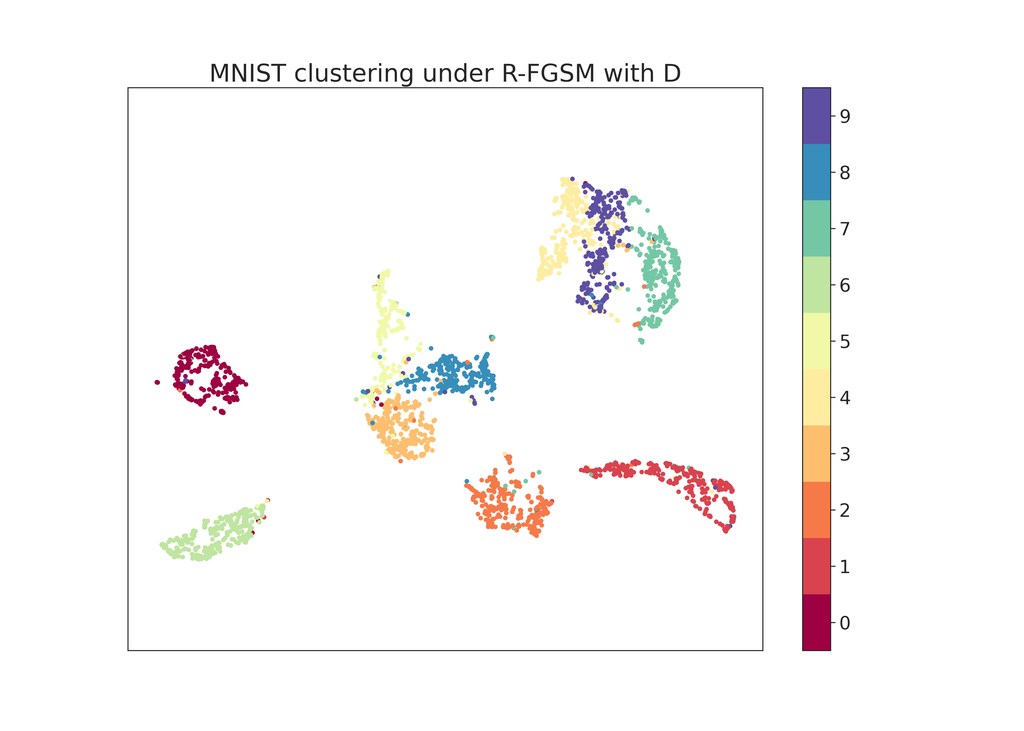}
  \caption{Rand-FGSM attacks}
  \label{fig:rfgsm-output-clustering}
\end{subfigure}
\begin{subfigure}{.33\textwidth}
  \centering
  \includegraphics[width=1.\linewidth]{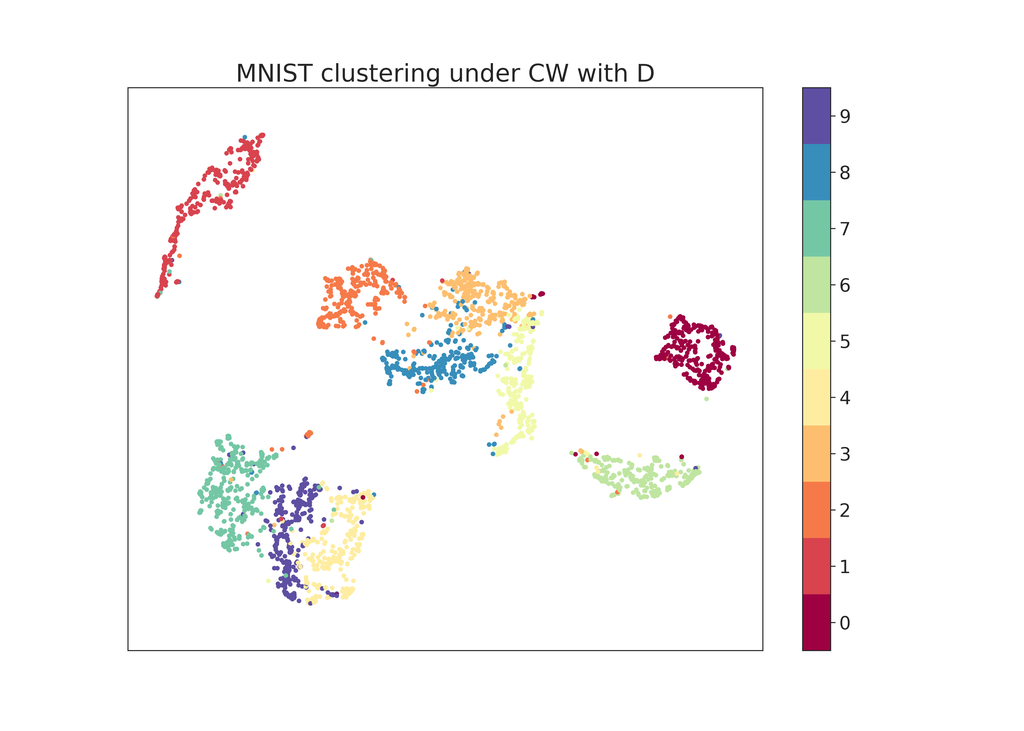}
  \caption{CW attack.}
  \label{fig:cw-output-clustering}
\end{subfigure}
\caption{Clustering plots of output from MAD-VAE with classification loss across different attack methods.}
\label{fig:output-clustering}
\end{figure}

Figure \ref{fig:output-clustering} shows the clustering result of the output from Defense-VAE with classification loss on adversarial data. The colored labels of Figure \ref{fig:output-clustering} are predicted using the classifier. From this figure, we are able to see that the output of our model is clearly clustered by color, whereas the adversarial input (Figure \ref{fig:clustering}) is not. Although some of the data with different labels are still mixed, it overall demonstrate the potential ability of the model to remap data back to the original data manifold.

\begin{figure}[H]
\begin{subfigure}{.33\textwidth}
  \centering
  \includegraphics[width=\linewidth]{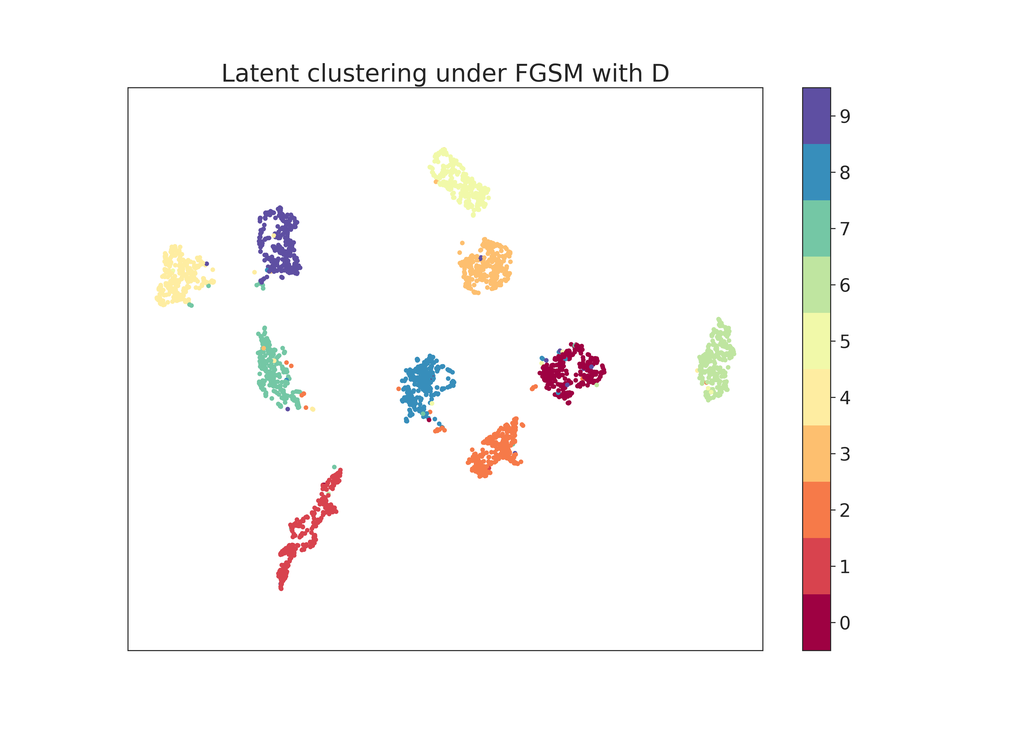}
  \caption{FGSM attack.}
  \label{fig:fgsm-latent-clustering}
\end{subfigure}%
\begin{subfigure}{.33\textwidth}
  \centering
  \includegraphics[width=\linewidth]{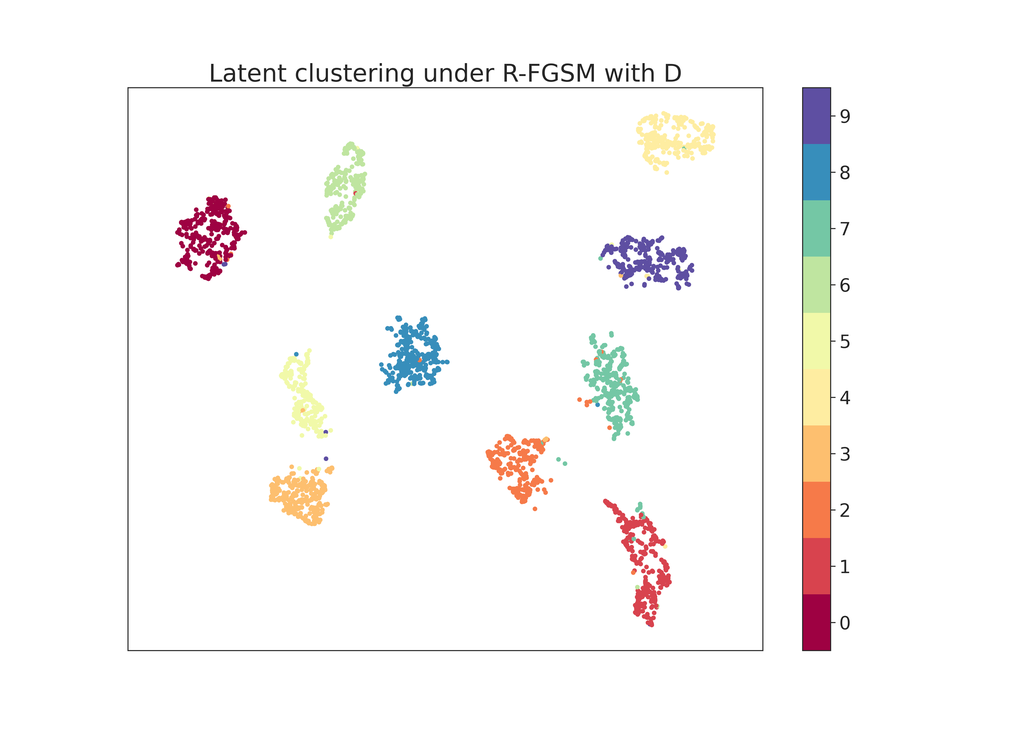}
  \caption{Rand-FGSM attacks}
  \label{fig:rfgsm-latent-clustering}
\end{subfigure}
\begin{subfigure}{.33\textwidth}
  \centering
  \includegraphics[width=\linewidth]{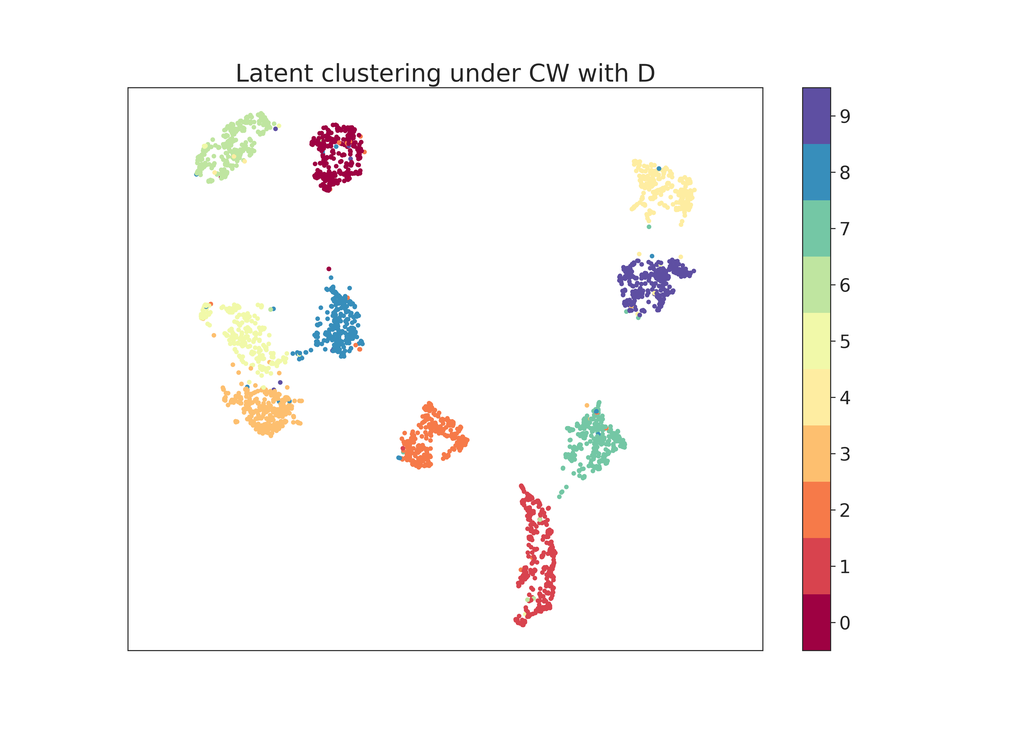}
  \caption{CW attack.}
  \label{fig:cw-latent-clustering}
\end{subfigure}
\caption{Clustering plots for latent variables of the MAD-VAE with classification loss given adversarial examples as input.}
\label{fig:latent-clustering}
\end{figure}

Figure \ref{fig:latent-clustering} shows the clustering of the latent variable $z$ in the classification MAD-VAE. Since we are adding the classification loss with respect to the latent variables in our new loss function, we would hope to see the clear separation of latent variables compared to the vanilla Defense-VAE.

\subsection{Clustering for Overpowered Attack} \label{sec:clustering-op}
For the clustering of Overpowered Attacks, we feel like it would be most representative of the attack to plot only one noise robustness set at a time. Therefore, in this section we only present the clustering for the Overpowered Attack with a noise robustness term of $1.0$. Additionally, our plots will focus on the Overpowered Attack against the proximity and distance loss MAD-VAE, since the effect of the adversary is most pronounced under this model.

\begin{figure}[H]
    \centering
    \includegraphics[width=.45\linewidth]{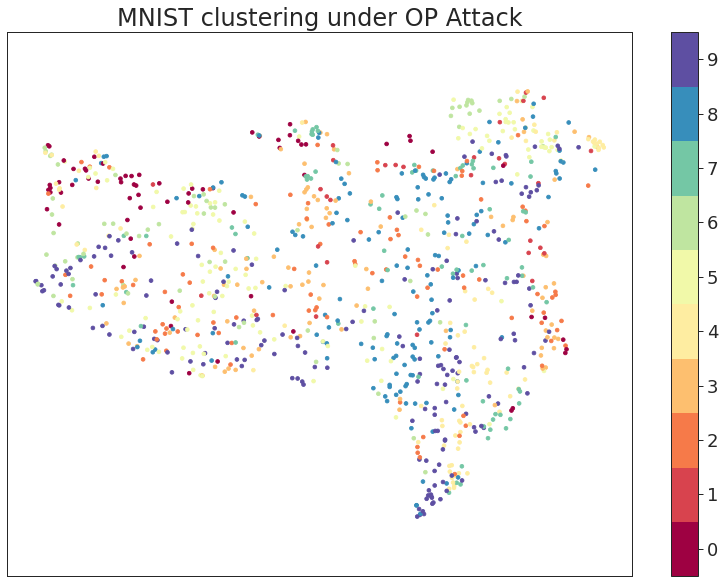}
    \caption{Adversarial clustering under the Overpowered Attack on MAD-VAE with proximity and distance loss.}
    \label{fig:mnist-op}
\end{figure}

Figure \ref{fig:mnist-op} shows the clustering of the output generated by the Overpowered Attack when attacking MAD-VAE with proximity and distance loss. In contrast to Figure \ref{fig:output-clustering}, we see that there is no obvious clustering structure associated with the underlying MNIST dataset. This is due to the fact that instead of attacking the ambient space of the classifier, the Overpowered Attack instead attacks the latent space of the defensive model. Intuitively, if the latent space accurately models the underlying data manifold of MNIST, then latent variables that lie outside of the learned manifold do not necessarily have to follow the underlying clustering when decoded.

\begin{figure}[H]
\begin{subfigure}{.5\textwidth}
  \centering
  \includegraphics[width=.75\linewidth]{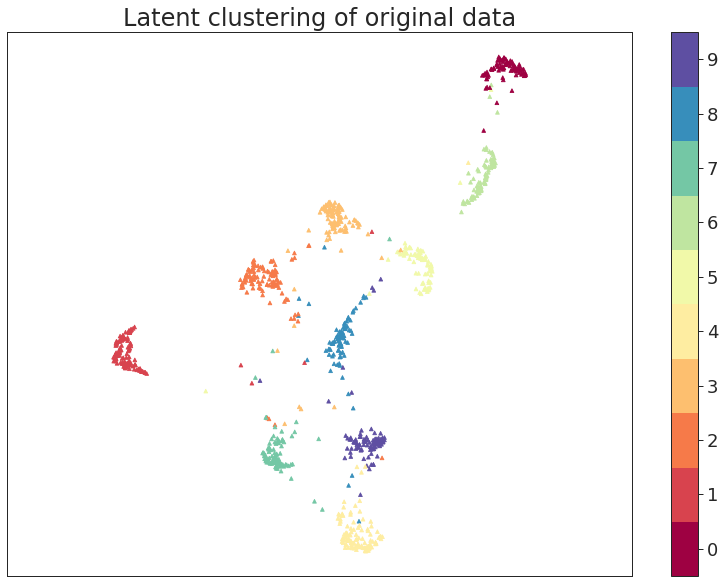}
  \caption{Unperturbed data clustering}
  \label{fig:op-orig}
\end{subfigure}%
\begin{subfigure}{.5\textwidth}
  \centering
  \includegraphics[width=.75\linewidth]{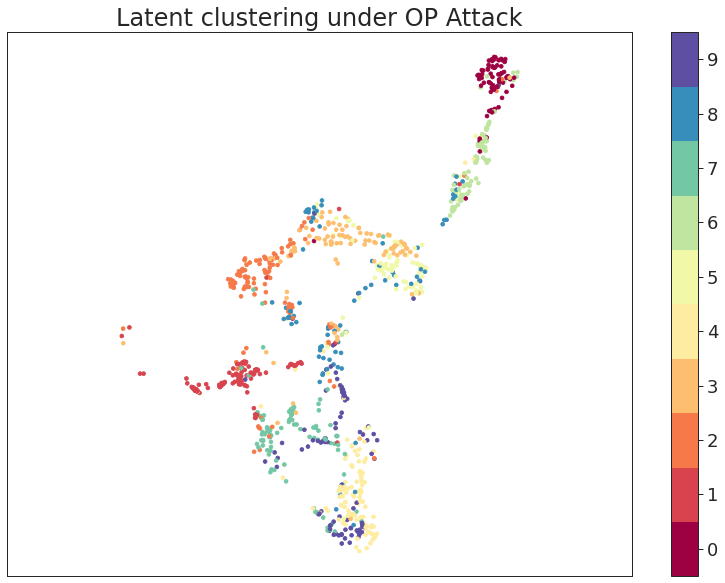}
  \caption{Overpowered Attack clustering}
  \label{fig:op-latent}
\end{subfigure}
\caption{Clustering plots for latent variables of the MAD-VAE with proximity and distance loss. First figure is original data and second under the Overpowered Attack.}
\label{fig:latent-op-clustering}
\end{figure}

In the next two plots, we will look at the effect of the Overpowered Attack on the latent variables of the defensive models. For clarity, we have split the original data and the adversarial data plots into two separate plots, though they should be viewed together. Figure \ref{fig:latent-op-clustering} shows the latent embedding of both the original data and the adversarial data generated by the Overpowered attack.

What we can see from Figure \ref{fig:latent-op-clustering} is that the Overpowered Attack tries to pull the original data points towards other clusters of data. The decoded images of the perturbed data are therefore more similar to other classes of numbers. The moving out data out of well-defined clusters towards other points could explain why Defense-VAE outperforms MAD-VAE in these tests, as Defense-VAE has a less structured latent space, as is shown in the Appendix Figure \ref{fig:vanilla-latent-clustering}. This, however, remains pure conjecture.

\subsection{Comments on the Overpowered Attack}
By exploiting the latent space of defensive models, the Overpowered Attack \cite{jalal2017robust} is very effective at finding what traditional literature on the subject considers adversarial examples. The traditional methodology surrounding adversarial examples is to find some adversarial perturbation under some $\ell_2$ distance budget from the original data point. It is natural to limit adversarial examples in this way since it prevents examples from becoming too noisy and defeating the ``imperceptible'' nature of adversarial examples.

\begin{figure}[H]
\begin{subfigure}{.5\textwidth}
  \centering
  \includegraphics[width=.5\linewidth]{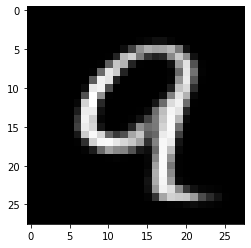}
  \caption{Unperturbed MNIST Data}
\end{subfigure}%
\begin{subfigure}{.5\textwidth}
  \centering
  \includegraphics[width=.5\linewidth]{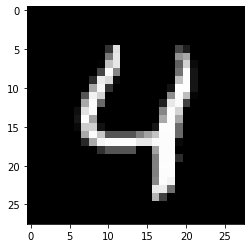}
  \caption{Overpowered Attack Data}
\end{subfigure}
\caption{Original and perturbed MNIST data under Overpowered Attack. $\ell_2$ difference of $3.8232$.}
\label{fig:op-abuse}
\end{figure}

While the methodology of limiting perturbations under some $\ell_2$ distance makes sense for ambient space attacks, for latent space attacks on defensive models, this methodology falls apart. The Overpowered Attack is one adversarial example generation method that seems to abuse this $\ell_2$ constraint. How does one abuse a distance metric? Well, latent space attacks abuse how inadequately the $\ell_2$ norm represents perceptual distance. The problem imposed by the $\ell_2$ norm surfaces as the adversarial example often being perceptually of a different class, although it originated from perturbing an image of a different class. Figure \ref{fig:op-abuse} demonstrates this failure of the $\ell_2$ norm, although there exists many more examples.

The implications of the phenomenon of latent space attacks transmuting an data point of one class to another raises some issues about how we measure the effectiveness of latent space attacks, and whether they can be used at all. In terms of creating defenses for latent space attacks, it is difficult to gauge accuracy since it is hard to know what images have changed human perceptible classes. Similarly, it is difficult to perform adversarial training (either strengthening a classifier or a defense model) because this attack may introduce incorrectly classified images as true. One possible solution may be to constrain the $\ell_2$ distance further, or perhaps to use a different metric altogether. I leave this problem of latent space class transmutation as an open problem to the reader.

\section{Conclusions and Future Works} \label{conclusion}
In this paper, we propose the MAD-VAE model which improves upon the robustness of Defense-VAE through three new training loss functions. While applied here to Defense-VAE, our loss function and methodology is not limited to any model and can be applied to any generative model based defense algorithms. We empirically show that our methods are more effective against various attack mechanisms than Defense-VAE.

The only attack method which MAD-VAE does not improve upon the accuracy of Defense-VAE is the Overpowered \cite{jalal2017robust} Attack introduced by Jalal et al. While empirically our method fails to outperform Defense-VAE, in this paper we have raised concerns as to the applicability of the Overpowered Attack and other latent space attacks in their current state. It is possible that our drop in accuracy is due to an increase of images in which they have been transmuted to a different class.

An aspect that worth delving into would be adapting our training methods to other models such as Defense-GAN. Since it is training a GAN is not an easy task, it was not feasible to complete it within the limited time-frame given to do this research. Additionally, the choice of different hyperparameters is also worth tuning for a longer period of time. In order to achieve better performance, a larger training data set could be generated and could be used to better tune the hyperparameters. Additionally, additional research on latent space attacks could shed more light on avoiding the class transmutation problem.

\section*{Acknowledgement}
We are thankful for the help and insightful suggestions offered by Professor Shuyang Ling at New York University Shanghai.

\newpage

\bibliographystyle{unsrt}  
\bibliography{references}  %

\newpage

\begin{appendices}
\section{Related Work and Background Information} \label{related}
We have based our research on Defense-VAE and made some improvements using different loss functions and training methods. Before talking in detail about our work, we will discuss different attack and defense algorithms first as well as introduce VAE and Defense-VAE in detail. Finally, we will mention the problems and drawbacks of current research.

\subsection{Attack Methods}
There are various attack algorithms haven been proposed. All of these attacks can be summarized in the following form: $\tilde{x} = x + \eta$ where $x$ is the original data and $\eta$ is the perturbations being added. Most of these attack methods aim to find the minimum perturbation $\eta$ which is undetectable by human beings but result in misclassification of the classifier.

\subsubsection{White-Box Attacks}
In the White-Box Attack scenario, the attackers will have full access to the classification model and the pretrained parameters. Therefore, the attackers can utilize the gradient of the model $J(\theta)$ to design attack algorithsm. Similar to the attack methods being used in Defense-GAN \cite{samangouei2018defense} and Defense-VAE \cite{defense-vae18}, we are using three major attacks: FGSM (fast gradient sign method) \cite{goodfellow2014explaining}, r-FGSM (random fast gradient sign method) \cite{tramr2017ensemble} and CW (Carlini-Wagner attack) \cite{carlini2016evaluating}.

\textbf{Fast Gradient Sign Method} is proposed by Ian Goodfellow \cite{goodfellow2014explaining}, where the attack is achieved by taking a single step in the direction of gradient of the classification model. The perturbation added to the image is $\eta = \epsilon sign(\nabla_x(\theta, x, y))$.

\textbf{Randomized Fast Gradient Sign Method} is an enhanced attack algorithm \cite{tramr2017ensemble} based on the FGSM attack. Instead of directly getting the gradient on the original image data $x$, r-FGSM added random noise on the image $x' = x + \alpha \cdot sign(N(0^d, I^d))$, and then doing the FGSM attack on the generated image $x'$: $x^{adv} = x' + (\epsilon - \alpha) \cdot sign(J_{x'}(\theta, x', y_{true}))$.

\textbf{Carlini-Wagner attack} is a powerful attack \cite{carlini2016evaluating} and it result in $0\%$ accuracy of the classifier under most of the circumstances. The attack is performed by solving the following optimization problems \cite{carlini2016evaluating} \cite{samangouei2018defense}:
\begin{align*}
    \min_{\delta \in R^n} & = ||\delta||_p + c \cdot f(x + \delta) \\
    s.t. & \hspace{0.5cm} x + \delta \in [0, 1]^n
\end{align*}
where $f$ is an objective function that will result in the misclassification of the model and $c$ is a constant.

\subsubsection{Black-Box Attacks}
Under the scenario of Black-Box attack, the attackers will have no access to the classification model. Therefore, it is often much more difficult for the attackers to perform successful and efficient attack on the model. In most of recent research works, two main methods have been proposed \cite{xu2019adversarial}: (1) find substitution models which will resemble the classifier, then use the the original White-Box attack methods\cite{papernot2016practical}, and (2) another method is to use efficient query methods by estimating the gradient from model output or repeatedly randomly adding certain perturbations \cite{guo2019simple}.

\subsection{Defenses Algorithms}
In addition to the release of numerous attack methods there has also been extensive research into defense mechanisms. Although there are many of them, we will only introduce two major algorithms in the following subsections.

\subsubsection{Adversarial Training}
One of the major ways to make classification model more robust against different adversarial attacks is to augment the training data set. The process of training classification model on a integrated data set of benign data and adversarial examples generated using different adversarial models is called adversarial training \cite{goodfellow2014explaining}. Although this method has been proved to be somewhat useful for defending attacks from attacks that are used to augment data, it is still vulnerable when facing a different attack that is not included in the combined training data set.

\subsubsection{Deep Generative Models and Denoising}
Another popular method is through leveraging the phenomenon of exploding and vanishing gradients in deep neural networks \cite{xu2019adversarial}. By first inputting the image into a generative model such as PixelDefend \cite{song2017pixeldefend} or Defense-GAN \cite{samangouei2018defense} before performing classification, the combined model is deeper and thus is subject to exploding and vanishing gradients. Additionally, the defense model is able to transfer the adversarial example back to the benign data manifold and thus, these generative defensive models can be viewed as a certain type of purifier. This is the method that is utilized in Defense-VAE \cite{defense-vae18}.

\subsection{Variational Autoencoders (VAE)}
Variational Autoencoders (VAE) are powerful generative models based on the ideas of Autoencoders and variational inference. The model consists of an encoder $Q(z|X)$ and a decoder $P(X|z)$. The encoder tries to encode the information of input data $X$ onto a latent variable $z$, while the decoder aims to completely reconstruct the data $X$ using the latent variable $z$.

In order to maker sure our encoder successfully approximate the posterior $P(z|X)$, we should calculate the Kullback–Leibler divergence (KL divergence or denoted by $KL$) between this two distributions.
\begin{align*}
    KL\left[Q(z|X)||P(z|X)\right] = E\left[\log(Q(z|X)) - \log(P(z|X))\right]
\end{align*}
Using Bayes' rule, we are able to get the objective function of VAE as follows:
\begin{align*}
    \log(P(X)) - KL\left[Q(z|X)||P(z|X)\right] = E\left[\log(P(X|z))\right] - KL\left[Q(z|X)||P(z)\right]
\end{align*}
We want to maximize the term on the left, since the KL divergence should be close to $0$ if we want our $z$ to be good enough to reproduce the data. The two terms on the right hand side ($ELBO$: evidence lower bound) can be optimized using gradient descent: the former one measures how well we reconstruct the original data, and the latter one bounds the latent $z$'s distribution to the real probability distribution $P(z)$ which we usually assume it to be standard normal distribution.

\subsection{Defense-VAE}
The original VAE model is not suitable to be used in adversarial defenses, since we do not want to reconstruct images still containing the information of perturbations. In order to solve this problem, Defense-VAE modifies the encoder and decoder as follows \cite{defense-vae18}:
\begin{align*}
    z \sim Enc(\tilde{X}) = Q(z|\tilde{X}), \hspace{0.5cm} X \sim Dec(z) = P(X|z)
\end{align*}
where $\tilde{X}=X+\eta$ is the adversarial example created by adding perturbation $\eta$ on the original data $X$. Since the distribution of the input data has been changed, the objective function of Defense-VAE changes as follows:
\begin{align*}
    ELBO = E_{Q(z|\tilde{X})}\left[\log(P(X|z))] - KL[Q(z|\tilde{X})||P(z)\right]
\end{align*}
where the input of Defense-VAE are the adversarial examples while the output are the corresponding benign data. 

The training process of Defense-VAE is very straight forward: we just need to use different attack algorithms to generate adversarial example based on the clean data thus creating a training pair. In reality, since we can use any attack methods with various parameters, we are able to generate large number of training data.

\subsection{Problems and Drawbacks of Current Research} \label{problem}
Despite the best efforts of the respective authors, the state-of-art gradient obfuscating defense algorithms such as Defense-GAN \cite{samangouei2018defense} and Defense-VAE \cite{defense-vae18} are still vulnerable to adversarial attacks \cite{jalal2017robust} \cite{athalye2018obfuscated}. Given a certain input image $X$, an attack can find another input $X'$ which is $\epsilon$-close to the original data $X$ that the classification output $C(X)$ and $C(X')$ are significantly different. Therefore, if such $(X, X')$ pair exists, then we can find a pair of $z$ accordingly: $(z, z')$ where $X\sim G(z)$ and $X' \sim G(z')$, yet $C(G(z))$ and $C(G(z'))$ are much different. Such attack method can be formulated as the following optimization problem:
\begin{align*}
    \sup_{z,z'} L(C(G(z)), C(G(z'))), \\
    s.t. ||G(z') - G(z)||_2^2 \leq (2\eta + \epsilon)^2
\end{align*}
where $L$ is some loss function, $\epsilon$ is the perturbation distance bound and $\eta$ is the distance bound between the generated data manifold $G$ and the real image manifold.

In their paper, Jalal et al. \cite{jalal2017robust} proposed the Overpowered Attack which is based on this formulation, resulting in decreasing the accuracy of Defense-GAN to only $3\%$. In addition, they also proposed the according defense methods: adversarial training with the Overpowered Attack, which makes the model powerful against PGD (projected gradient descent attack, another popular White-Box attack) \cite{aleks2017deep}.

There are a couple advantages to the approach proposed by Jalal et al. Firstly, by searching for adversarial pairs in the typically much lower-dimensional latent space of the defensive model, the optimization problem of maximizing the classifier loss is made significantly easier. Secondly, due to searching in the latent space of the defensive model as opposed to the ambient space, the adversarial images that are captured are endowed with much more relevant information as opposed to noise.

\subsection{Topological Structures in Adversarial Defense} \label{topological}
In an effort to formalize the problem of adversarial examples while using defensive models, Jang et al. \cite{jang2019need} suggest a underlying cause for adversarial examples and offer up a solution. Jang et al. proposes that the existence of such adversarial examples stems from the lack of understanding of the topological structure of the underlying data manifold. Simply put, the learned data distribution does not match the true data distribution, and thus there exists images that do not get mapped back to the data manifold by the defense model.

Jang et al. prove a theorem which states that if the latent distribution is composed of $n_Z$ multivariate Gaussian distributions and the data manifold is composed of $n_X$ components, if $n_Z < n_X$ then there exists points outside the data manifold that are part of the approximated data manifold of the latent dimensions. In our proposed model, we use some of the methods they introduced in the October 2019 version of their paper in order endow our model with better topological awareness. Since then, the paper by Jang et al. has undergone some drastic changes, which would be worth looking at in the future.

\section{Neural Network Architectures}
The details of the models used in our paper and experiments are listed in the following three tables.

\begin{table}[H]
\centering
\begin{tabular}{|l|l|}
\hline
\textbf{Encoder} & \textbf{Decoder} \\ \hline
Conv(*, 64, 5, 1, 2) + BN + ReLU  & FC(128, 4096) + ReLU \\
Conv(64, 64, 4, 2, 3) + BN + ReLU & ConvT(256, 128, 4, 2, 1) + BN + ReLU \\
Conv(64, 128, 4, 2, 1) + BN + ReLU & ConvT(128, 64, 4, 2, 1) + BN + ReLU \\
Conv(128, 256, 4, 2, 1) + BN + ReLU & ConvT(64, 64, 4, 2, 3) + BN + ReLU \\
FC1(4096, 128), FC2(4096, 128) & ConvT(64, 64, 5, 1, 2) + BN + ReLU \\ \hline
\end{tabular}
\caption{Model structure for Defense-VAE \protect\cite{defense-vae18}}
\label{table:defense-vae}
\end{table}

\begin{table}[H]
\centering
\begin{tabular}{|l|l|}
\hline
\textbf{Layers} & \textbf{Params} \\ \hline
Conv.ReLU  &  3 $\times$ 3 $\times$ 32 \\
Conv.ReLU &  3 $\times$ 3 $\times$ 32 \\
Max Pooling & 2 $\times$ 2 \\
Conv.ReLU & 3 $\times$ 3 $\times$ 64 \\
Conv.ReLU & 3 $\times$ 3 $\times$ 64 \\
Max Pooling & 2 $\times$ 2 \\ 
Dense.ReLU & 200 \\
Dense.ReLU & 200 \\
Softmax & 10 \\
\hline
\end{tabular}
\caption{Model structure for classifier \protect\cite{meng2017magnet}}
\label{table:classifier}
\end{table}

\begin{table}[H]
\centering
\begin{tabular}{|l|l|l|l|}
\hline
\textbf{A} & \textbf{B} & \textbf{C} & \textbf{D, E*} \\ \hline
 Conv(64, 5 $\times$ 5, 1) & Dropout(0.2) & Conv(128, 3 $\times$ 3, 1) & FC(200)\\
 ReLU & Conv(64, 8 $\times$ 8, 2) & ReLU & ReLU \\
 Conv(64, 5 $\times$ 5, 2) & ReLU & Conv(64, 3 $\times$ 3, 2) & Dropout(0.5)\\
 ReLU  & Conv(128, 6 $\times$ 6, 2) & ReLU & FC(200) \\
 Dropout(0.25) & ReLU & Dropout(0.25) & ReLU\\
 FC(128) & Conv(128, 5 $\times$ 5, 1) & FC(128) & Dropout(0.5)\\ 
 ReLU & ReLU & ReLU & FC(10) + Softmax\\
 Dropout(0.5) & Dropout(0.5) & Dropout(0.5) &\\
 FC(10) + Softmax & FC(10) + Softmax & FC(10) + Softmax &\\
\hline
\end{tabular}
\caption{Model structure for Black-Box attack classifier \protect\cite{samangouei2018defense}. * indicates same structure without dropout layer.}
\label{table:black-box}
\end{table}

\section{Additional Plots for Experiments}
In this section, we will provide the clustering plots of model output and latent variables that have not been included in the sections above.

\begin{figure}[H]
\begin{subfigure}{.33\textwidth}
  \centering
  \includegraphics[width=\linewidth]{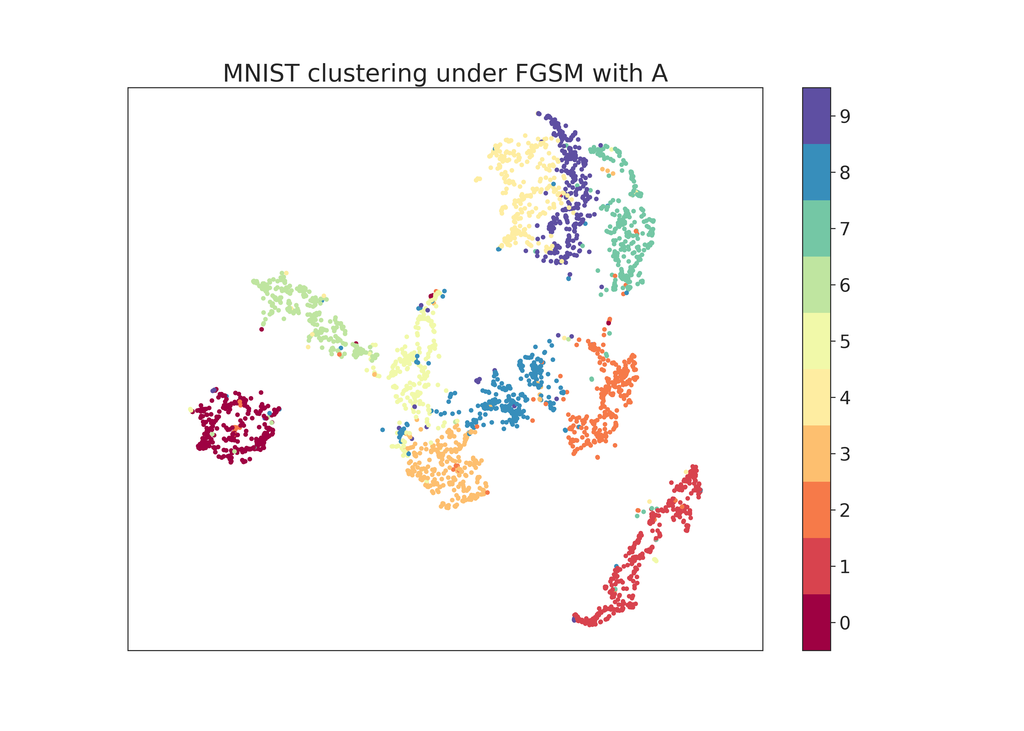}
  \caption{FGSM attack.}
  \label{fig:vanilla-fgsm-output-clustering}
\end{subfigure}%
\begin{subfigure}{.33\textwidth}
  \centering
  \includegraphics[width=\linewidth]{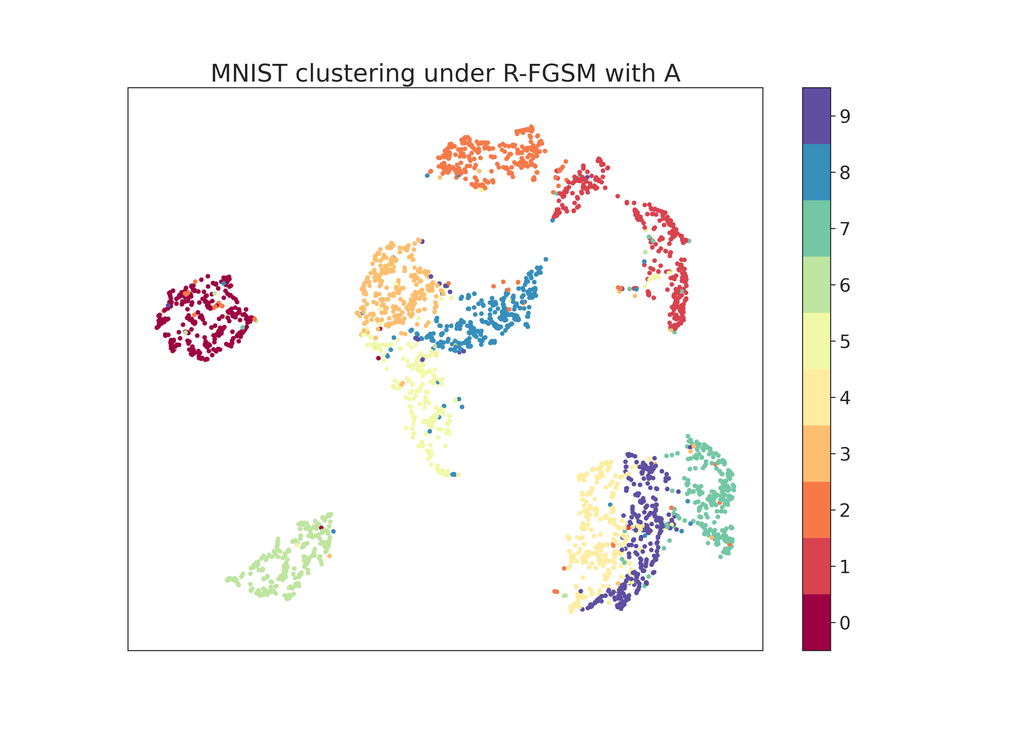}
  \caption{Rand-FGSM attacks}
  \label{fig:vanilla-rfgsm-output-clustering}
\end{subfigure}
\begin{subfigure}{.33\textwidth}
  \centering
  \includegraphics[width=\linewidth]{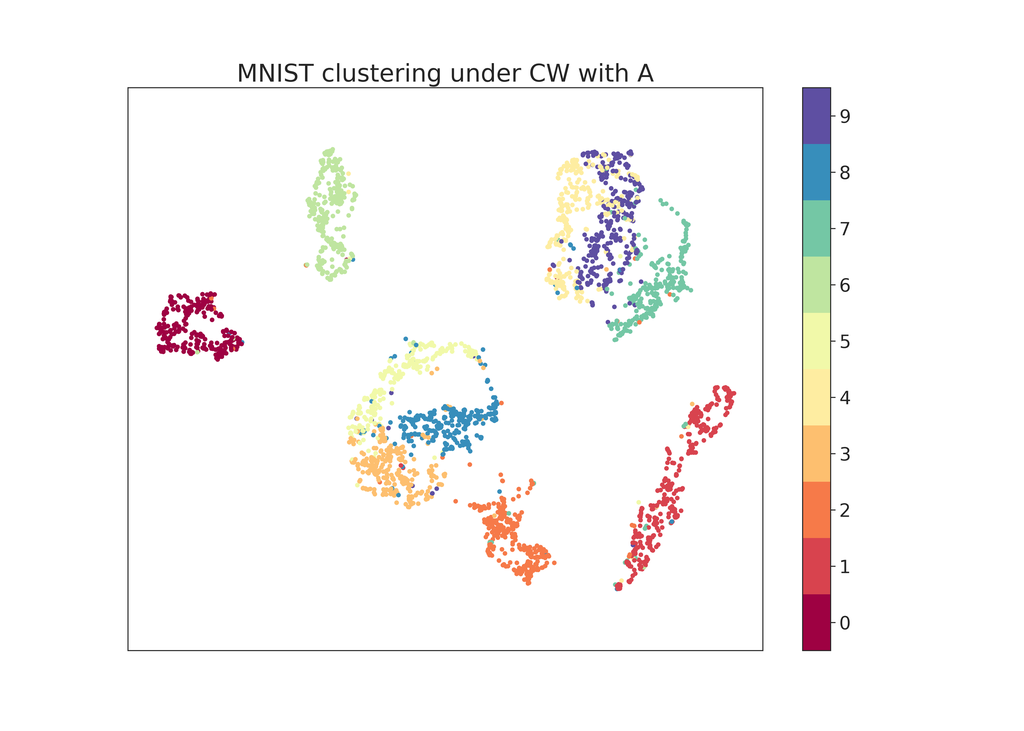}
  \caption{CW attack.}
  \label{fig:vanilla-cw-output-clustering}
\end{subfigure}
\caption{Clustering plots of output from vanilla model across various adversarial inputs.}
\label{fig:vanilla-output-clustering}
\end{figure}

\begin{figure}[H]
\begin{subfigure}{.33\textwidth}
  \centering
  \includegraphics[width=\linewidth]{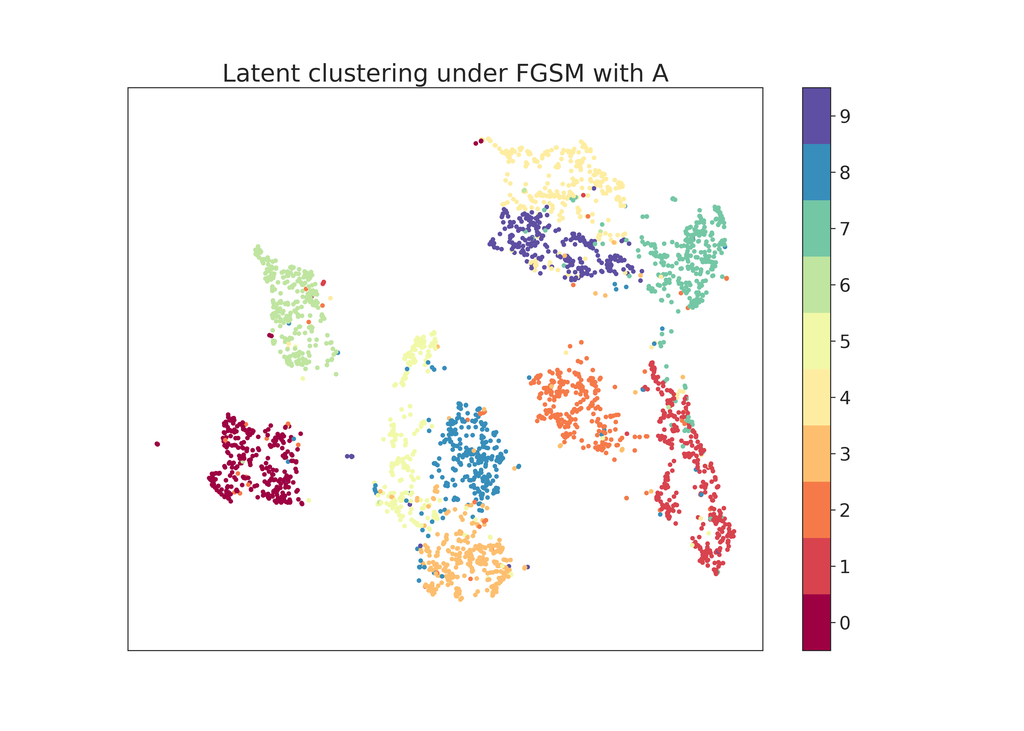}
  \caption{FGSM attack.}
  \label{fig:vanilla-fgsm-latent-clustering}
\end{subfigure}%
\begin{subfigure}{.33\textwidth}
  \centering
  \includegraphics[width=\linewidth]{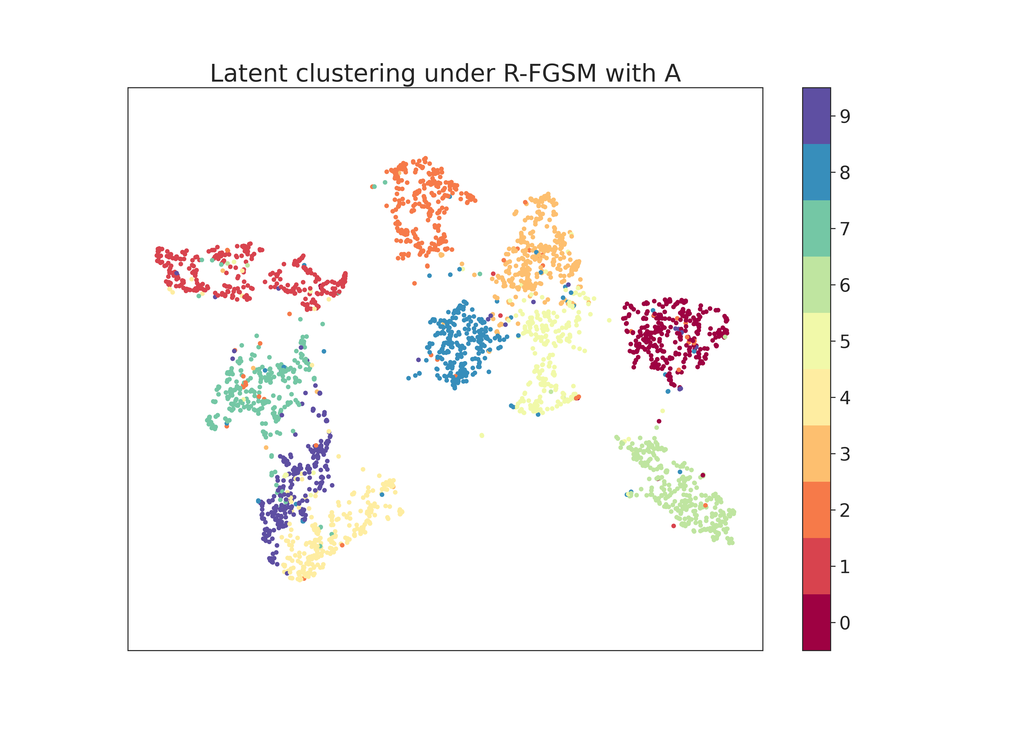}
  \caption{Rand-FGSM attacks}
  \label{fig:vanilla-rfgsm-latent-clustering}
\end{subfigure}
\begin{subfigure}{.33\textwidth}
  \centering
  \includegraphics[width=\linewidth]{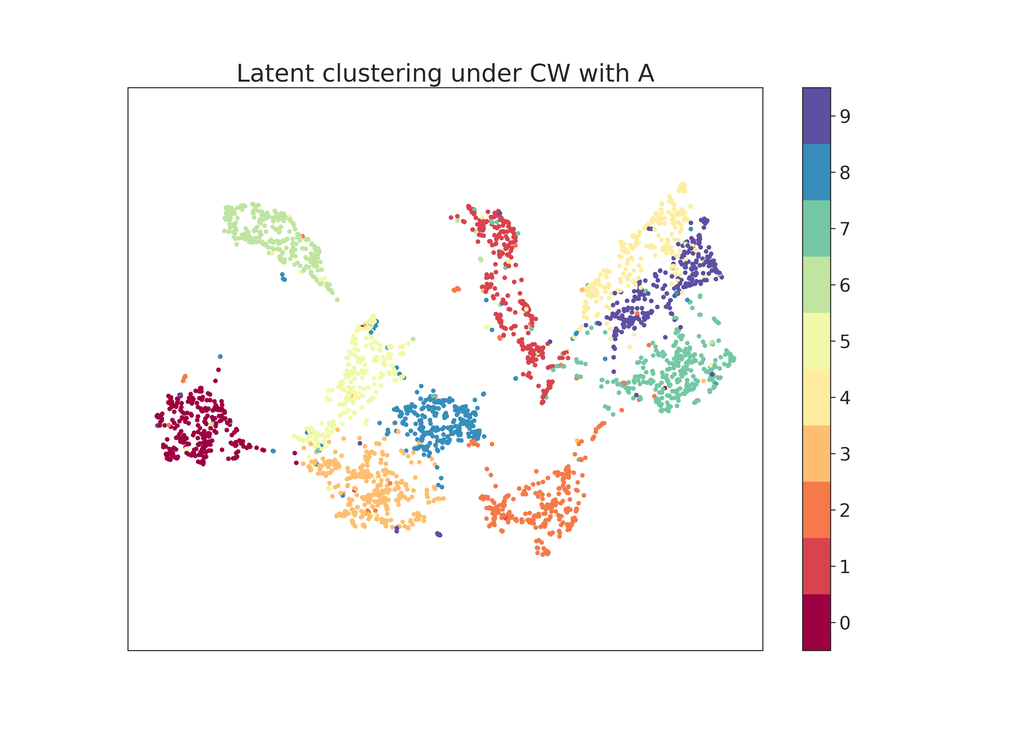}
  \caption{CW attack.}
  \label{fig:vanilla-cw-latent-clustering}
\end{subfigure}
\caption{Clustering plots for latent variables of the vanilla model given adversarial examples as input.}
\label{fig:vanilla-latent-clustering}
\end{figure}

\begin{figure}[H]
\begin{subfigure}{.33\textwidth}
  \centering
  \includegraphics[width=1.\linewidth]{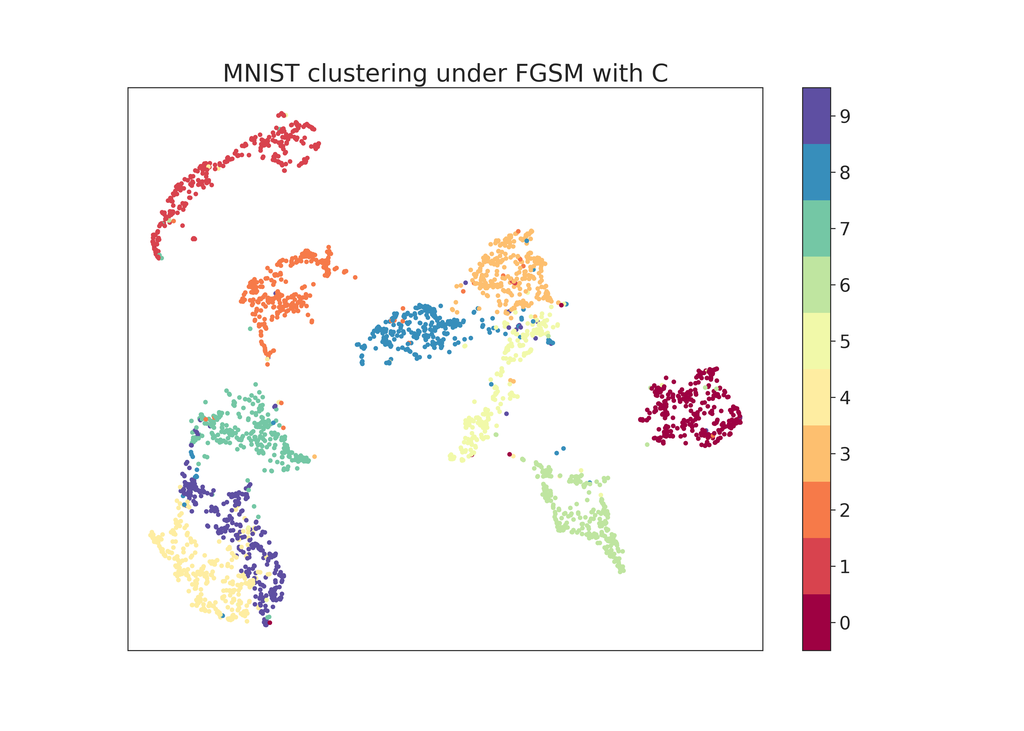}
  \caption{FGSM attack.}
  \label{fig:proxi-dist-fgsm-output-clustering}
\end{subfigure}%
\begin{subfigure}{.33\textwidth}
  \centering
  \includegraphics[width=1.\linewidth]{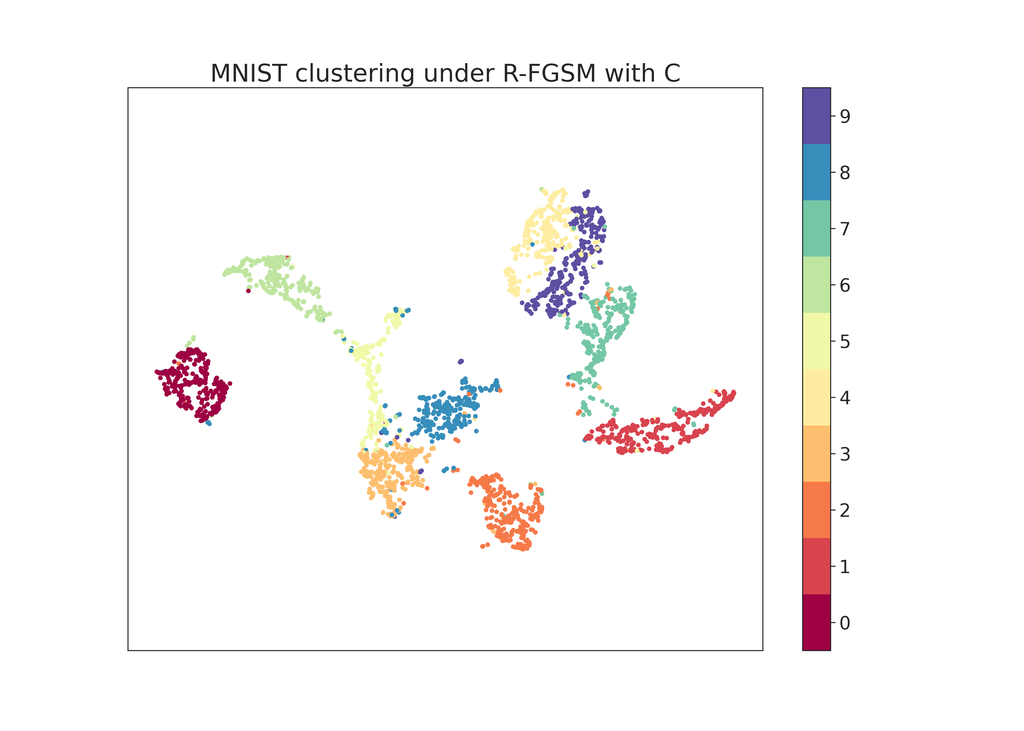}
  \caption{Rand-FGSM attacks}
  \label{fig:proxi-dist-rfgsm-output-clustering}
\end{subfigure}
\begin{subfigure}{.33\textwidth}
  \centering
  \includegraphics[width=1.\linewidth]{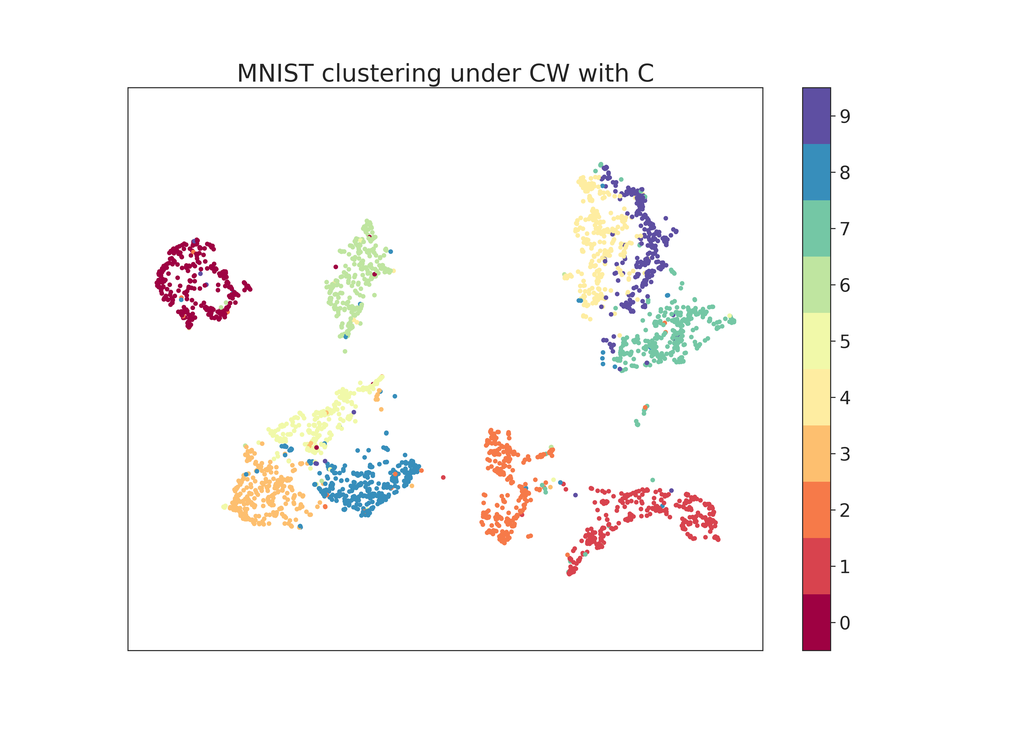}
  \caption{CW attack.}
  \label{fig:proxi-dist-cw-output-clustering}
\end{subfigure}
\caption{Clustering plots of output from model trained with proximity and distance loss across various adversarial inputs.}
\label{fig:proxi-dist-output-clustering}
\end{figure}

\begin{figure}[H]
\begin{subfigure}{.33\textwidth}
  \centering
  \includegraphics[width=\linewidth]{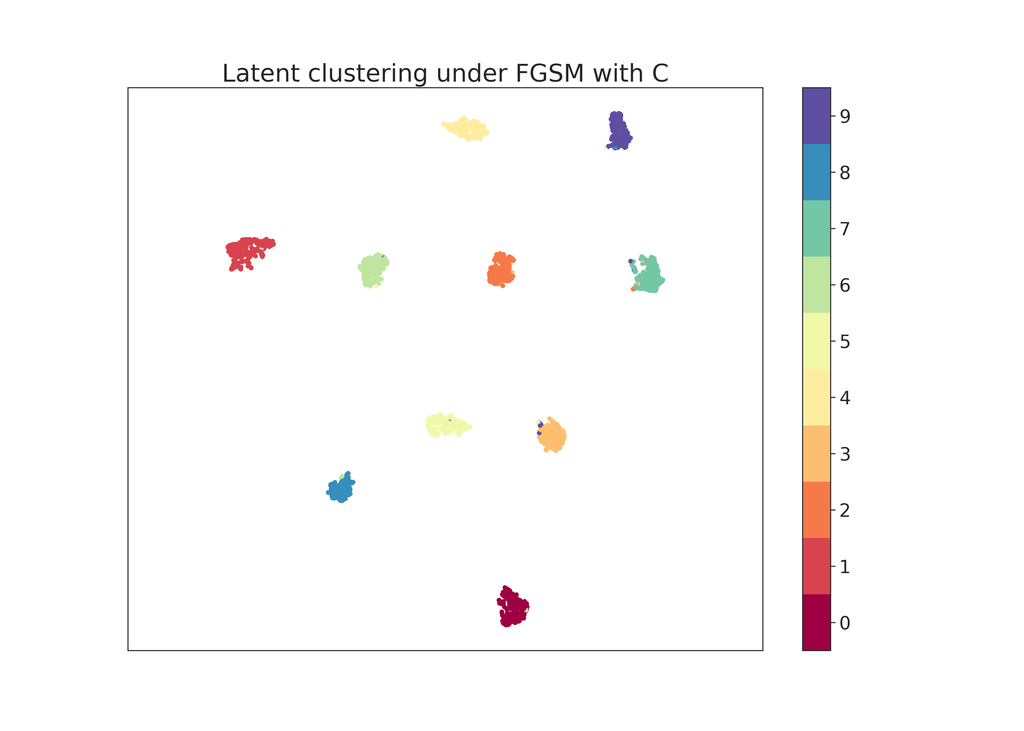}
  \caption{FGSM attack.}
  \label{fig:proxi_dist-fgsm-latent-clustering}
\end{subfigure}%
\begin{subfigure}{.33\textwidth}
  \centering
  \includegraphics[width=\linewidth]{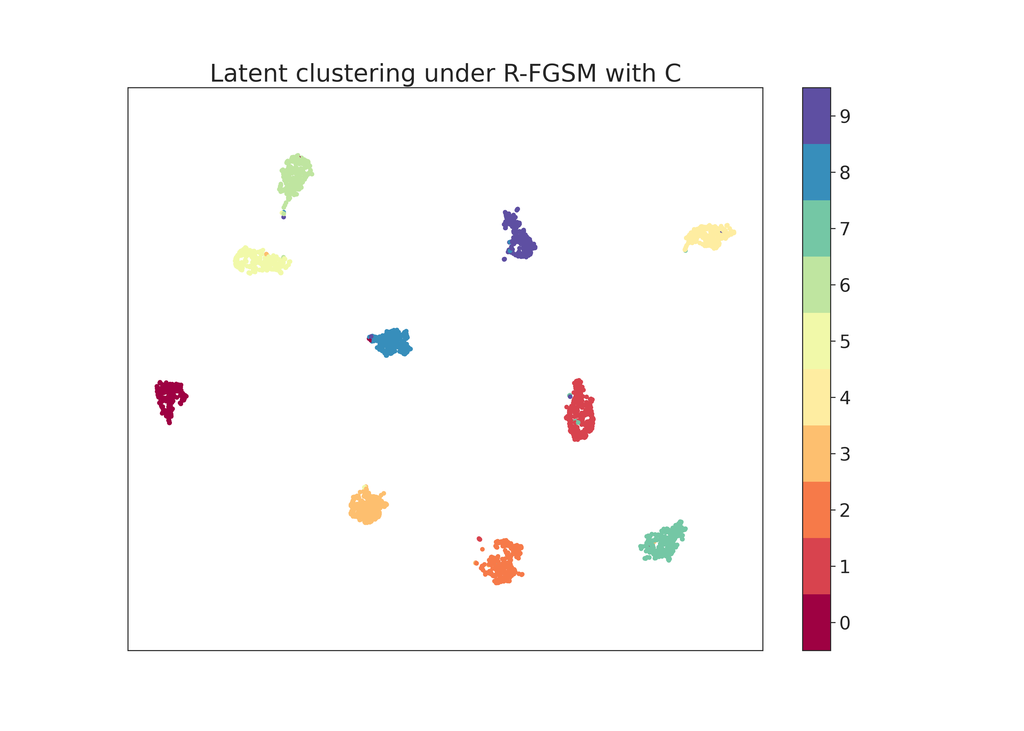}
  \caption{Rand-FGSM attacks}
  \label{fig:proxi_dist-rfgsm-latent-clustering}
\end{subfigure}
\begin{subfigure}{.33\textwidth}
  \centering
  \includegraphics[width=\linewidth]{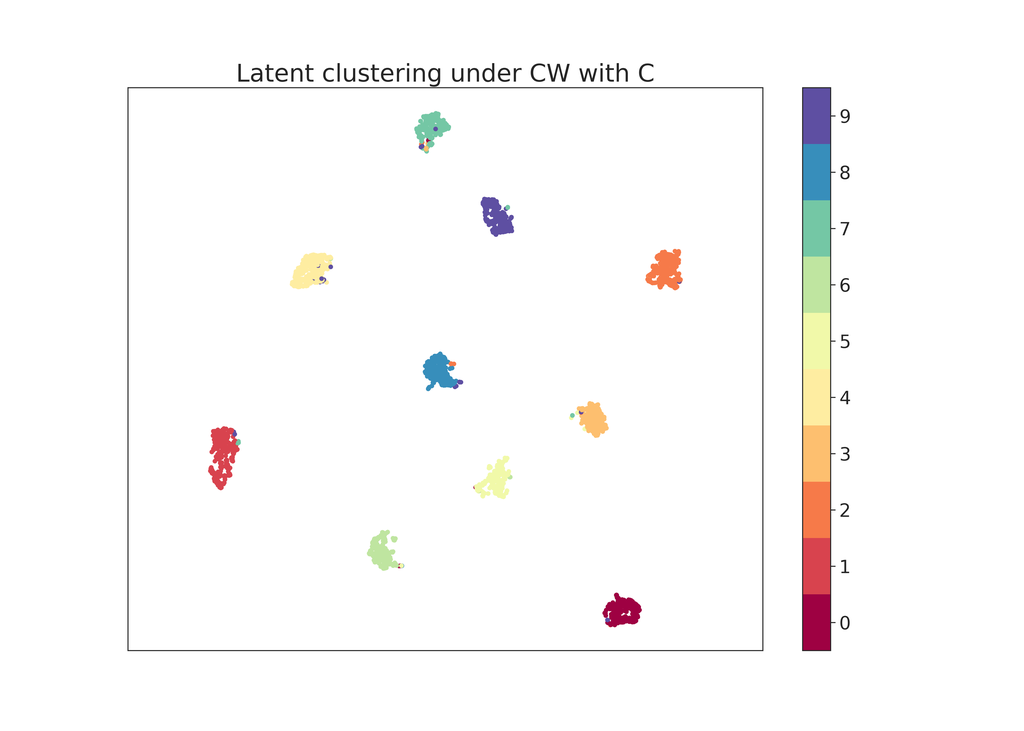}
  \caption{CW attack.}
  \label{fig:proxi_dist-cw-latent-clustering}
\end{subfigure}
\caption{Clustering plots for latent variables of the model trained with proximity and distance loss given adversarial examples as input.}
\label{fig:proxi_dist-latent-clustering}
\end{figure}

\begin{figure}[H]
\begin{subfigure}{.33\textwidth}
  \centering
  \includegraphics[width=1.\linewidth]{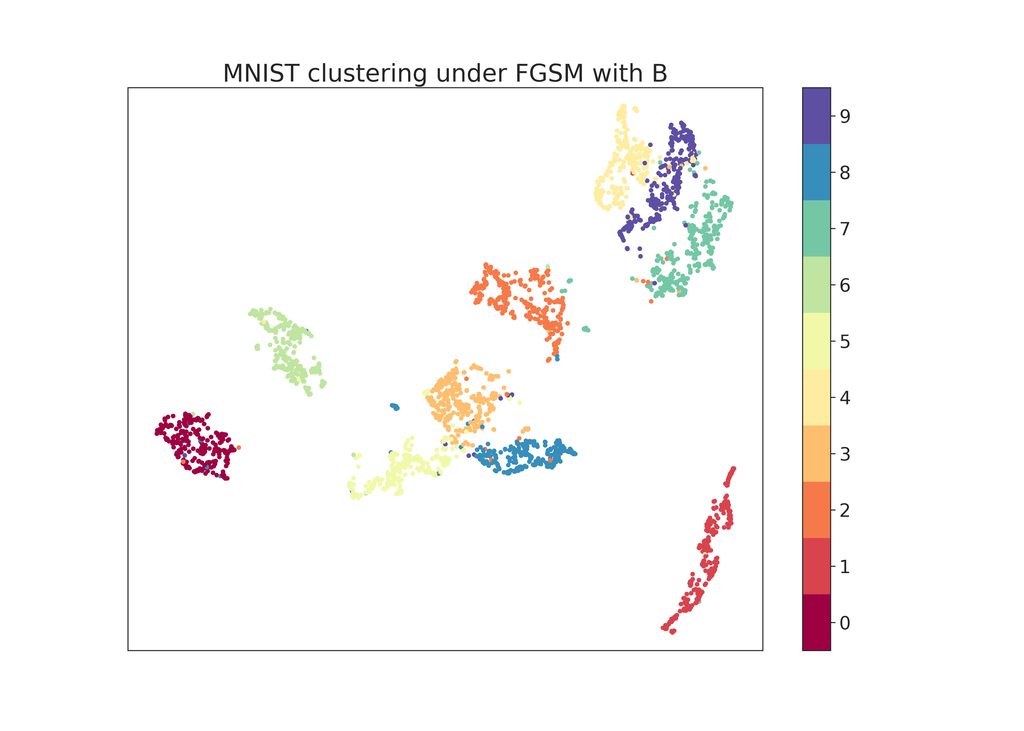}
  \caption{FGSM attack.}
  \label{fig:combined-fgsm-output-clustering}
\end{subfigure}%
\begin{subfigure}{.33\textwidth}
  \centering
  \includegraphics[width=1.\linewidth]{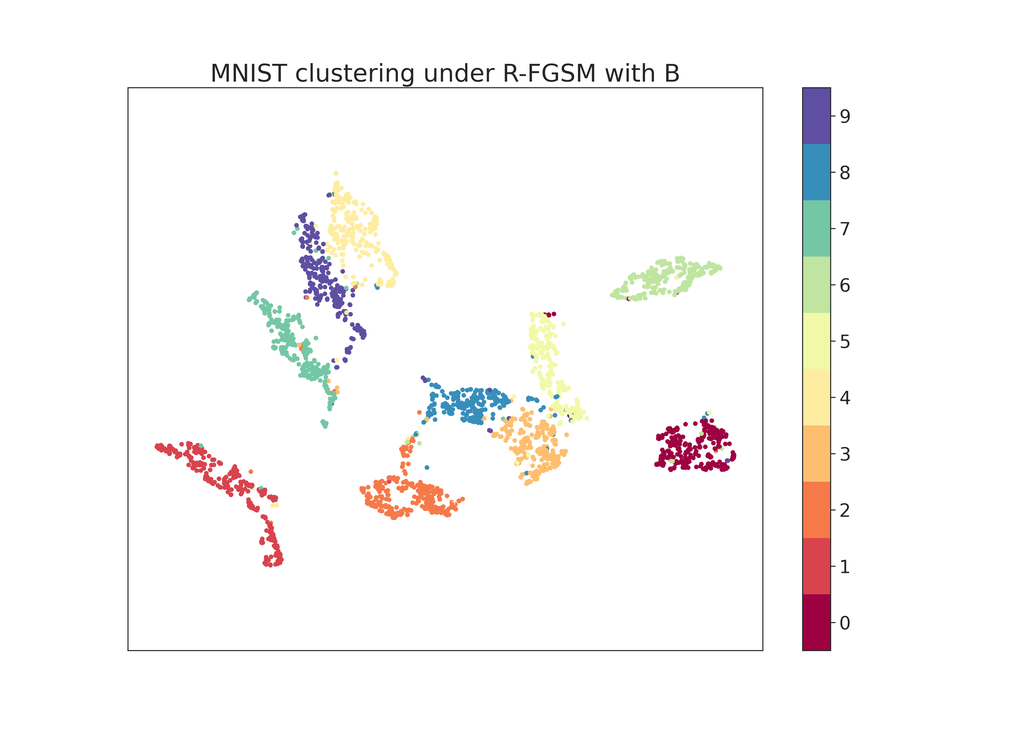}
  \caption{Rand-FGSM attacks}
  \label{fig:combined-rfgsm-output-clustering}
\end{subfigure}
\begin{subfigure}{.33\textwidth}
  \centering
  \includegraphics[width=1.\linewidth]{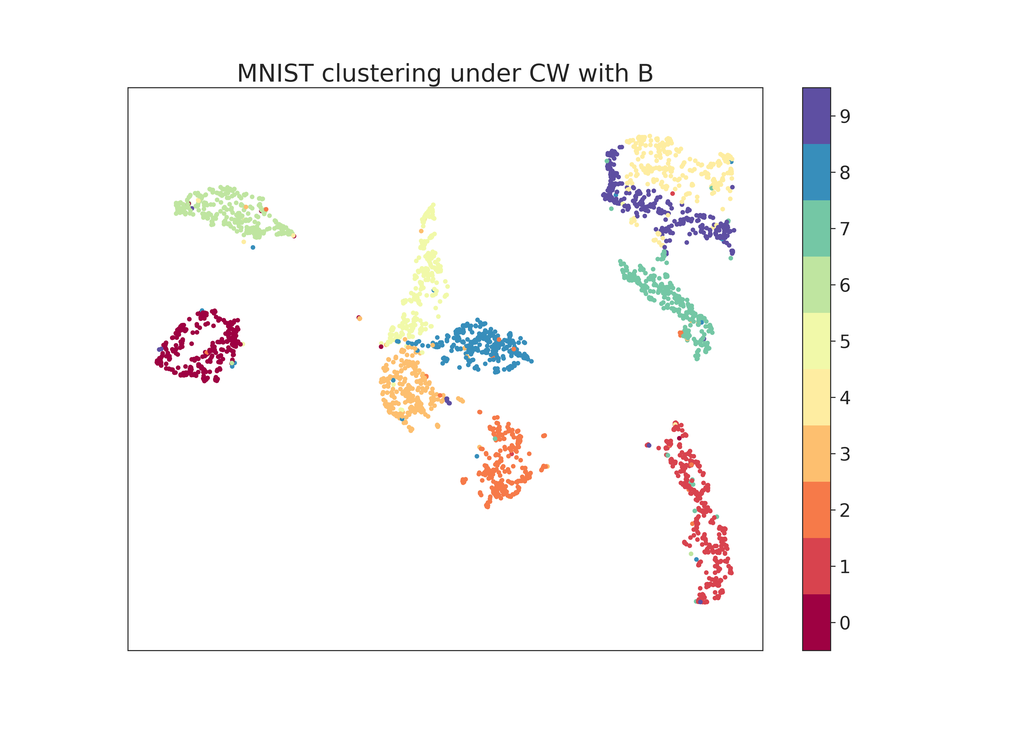}
  \caption{CW attack.}
  \label{fig:combined-cw-output-clustering}
\end{subfigure}
\caption{Clustering plots of output from model trained with combined loss across various adversarial inputs.}
\label{fig:combined-output-clustering}
\end{figure}

\begin{figure}[H]
\begin{subfigure}{.33\textwidth}
  \centering
  \includegraphics[width=\linewidth]{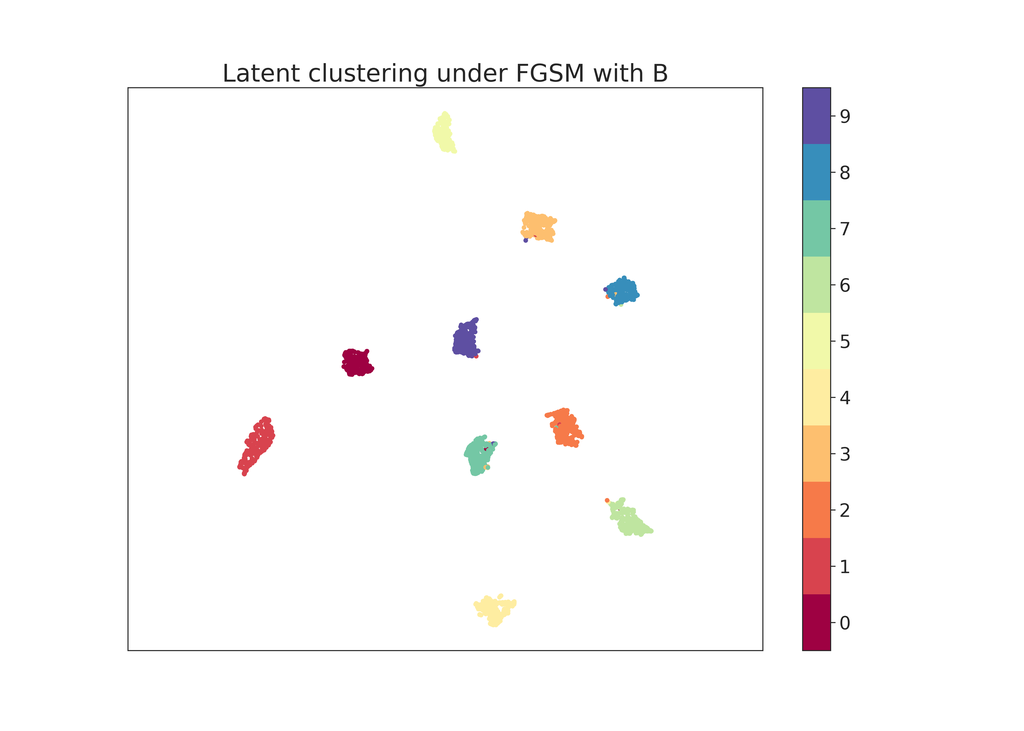}
  \caption{FGSM attack.}
  \label{fig:combined-fgsm-latent-clustering}
\end{subfigure}%
\begin{subfigure}{.33\textwidth}
  \centering
  \includegraphics[width=\linewidth]{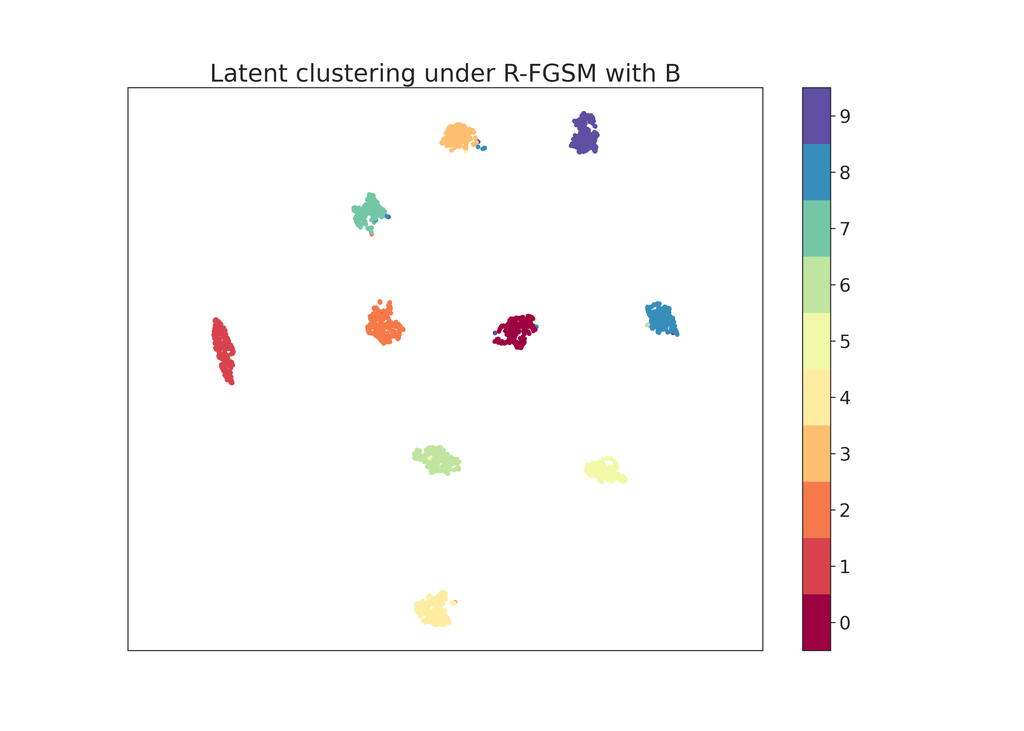}
  \caption{Rand-FGSM attacks}
  \label{fig:combined-rfgsm-latent-clustering}
\end{subfigure}
\begin{subfigure}{.33\textwidth}
  \centering
  \includegraphics[width=\linewidth]{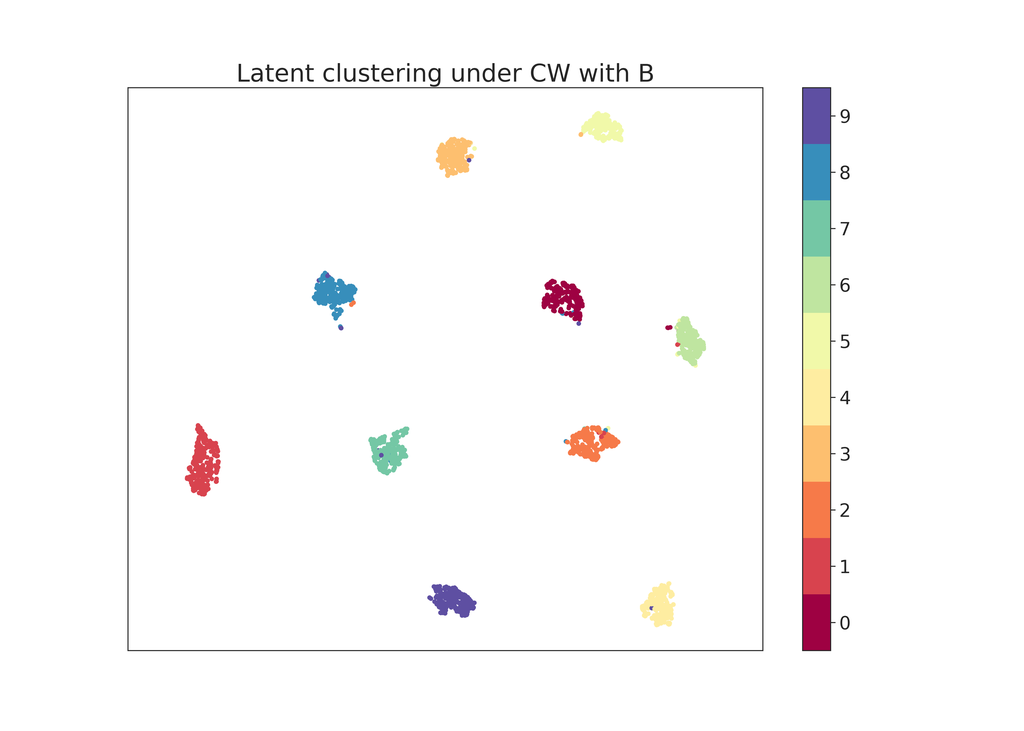}
  \caption{CW attack.}
  \label{fig:combined-cw-latent-clustering}
\end{subfigure}
\caption{Clustering plots for latent variables of the model trained with combined loss given adversarial examples as input.}
\label{fig:combined-latent-clustering}
\end{figure}

\end{appendices}

\end{document}

%% file: whitebox_table.tex
\begin{table}[H]
\centering
\begin{tabular}{c|c|c|c|c|c|c}
Attack & No Attack & No Defense & Vanilla & Classification & Proximity and Distance & Combined \\ \hline
FGSM & 0.9931 & 0.1316 & 0.9675 & 0.9899 & 0.9836 & \textbf{0.9905}\\
Rand-FGSM & 0.9931 & 0.1521 & 0.9734 & \textbf{0.9914} & 0.9862 & 0.99\\
CW & 0.9931 & 0.0075 & 0.9772 & 0.9906 & 0.985 & \textbf{0.9914}\\
MI-FGSM & 0.9931 & 0.0074 & 0.963 & \textbf{0.989} & 0.9816 & 0.9887\\
PGD & 0.9931 & 0.0073 & 0.9725 & \textbf{0.9883} & 0.9827 & 0.988\\
Single Pixel & 0.9931 & 0.9977 & 0.9845 & \textbf{0.9887} & 0.9877 & 0.9887\\
\end{tabular}
\caption{Classification accuracy of different models under different attacks with the default parameters. The models are trained on the data generated using the first three attack methods while the other three attacks are not included in the training dataset. Highest accuracy per attack is in bold.}
\label{table:whitebox-result}
\end{table}

%% file: blackbox_table.tex
\begin{table}[H]
\centering
\begin{tabular}{c|c|c|c|c|c|c}
Substitute & No Attack & No Defense & Vanilla & Classification & Proximity and Distance & Combined \\ \hline
A & 0.9939 & 0.0487 & 0.9615 & \textbf{0.9792} & 0.9698 & 0.9774\\
B & 0.9925 & 0.0144 & 0.9513 & \textbf{0.9726} & 0.9632 & 0.9704\\
C & 0.9938 & 0.0546 & 0.9653 & \textbf{0.9741} & 0.967 & 0.966\\
D & 0.9812 & 0.0173 & 0.9319 & 0.9419 & \textbf{0.95} & 0.9387\\
E & 0.9807 & 0.0155 & 0.9305 & 0.946 & \textbf{0.9528} & 0.9389\\
\end{tabular}
\caption{Classification accuracy of different models based on the FGSM Black-Box attack on various substitute models with $\epsilon=0.3$. Highest accuracy per substitute is in bold.}
\label{table:blackbox-result}
\end{table}

%% file: OP_table.tex
\begin{table}[H]
\centering
\begin{tabular}{c|c|c|c|c|c}
Metric & Identity & Vanilla & Classification & Proximity and Distance & Combined \\ \hline
Accuracy & 0.3078 & \textbf{0.4443} & 0.386 & 0.4345 & 0.3885\\
Original Accuracy & 0.199 & \textbf{0.335} & 0.273 & 0.329 & 0.284\\
\end{tabular}
\caption{Classification accuracy of different defensive models under the Overpowered Attack. The meaning of the two metrics is explained above. Highest accuracy per metric is in bold.}
\label{table:op-result}
\end{table}